\documentclass[fleqn,usenatbib]{mnras}

\usepackage{newtxtext,newtxmath}

\usepackage[T1]{fontenc}
\usepackage{ae,aecompl}

\usepackage{hyperref}
\usepackage{graphicx}	
\usepackage{blindtext}  
\usepackage{natbib}
\usepackage{rotating,times,graphicx,latexsym}
\usepackage{color}
\usepackage{longtable}
\usepackage{lscape}
\usepackage{lipsum} 
\usepackage{array}
\usepackage{flafter}

\usepackage{soul} 
\usepackage{xargs} 
\usepackage[pdftex,dvipsnames]{xcolor}  
\usepackage[colorinlistoftodos,prependcaption]{todonotes}
\newcommandx{\tobedone}[2][1=]{\todo[linecolor=red,backgroundcolor=red!25,bordercolor=red,inline,#1]{#2}}
\newcommandx{\changed}[2][1=]{\todo[linecolor=blue,backgroundcolor=blue!25,bordercolor=blue,inline,#1]{#2}\noindent}
\newcommandx{\thiswillnotshow}[2][1=]{\todo[disable,#1]{#2}}
\newcommandx{\carys}[2][1=]{\todo[linecolor=Orchid,backgroundcolor=Orchid!25,bordercolor=Orchid,inline,#1]{#2}\noindent}
\newcommandx{\dirk}[2][1=]{\todo[linecolor=BurntOrange,backgroundcolor=BurntOrange!25,bordercolor=BurntOrange,inline,#1]{#2}\noindent}
\newcommandx{\jcw}[2][1=]{\todo[linecolor=SpringGreen,backgroundcolor=SpringGreen!40,bordercolor=SpringGreen,inline,#1]{#2}\noindent}

\graphicspath{ {images_pdf/} }
\newcommand{\hii}{H{\sc ii}~}

\title[Spot properties on YSOs]{A survey for variable young stars with small telescopes: VII - Spot Properties on YSOs in IC\,5070}

\author[Carys Herbert et al.]{Carys Herbert$^{1}$\thanks{E-mail: cbah2@kent.ac.uk}, 
Dirk Froebrich$^{2}$,
Aleks Scholz$^{3}$
\\
$^{1}$Centre for Astrophysics and Planetary Science, School of Physical Sciences, University of Kent, Canterbury CT2 7NH, UK\\
$^{2}$School of Physical Sciences, University of Kent, Canterbury CT2 7NH, UK\\
$^{3}$SUPA, School of Physics \& Astronomy, University of St Andrews, North Haugh, St Andrews KY16 9SS, UK\\
}

\date{Accepted XXX. Received YYY; in original form ZZZ}

\pubyear{2022}

\begin{document}
\label{firstpage}
\pagerange{\pageref{firstpage}--\pageref{lastpage}}
\maketitle

\begin{abstract} We present measurements of spot properties on 31 young stellar objects, based on multi-band data from the HOYS (Hunting Outbursting Young Stars) project. On average the analysis for each object is based on 270 data points during 80 days in at least 3 bands. All the young low-mass stars in our sample show periodic photometric variations. We determine spot temperatures and coverage by comparing the measured photometric amplitudes in optical bands with simulated amplitudes based on atmosphere models, including a complete error propagation. 21 objects in our sample feature cool spots, with spot temperatures 500\,--\,2500\,K below the stellar effective temperature ($T_{\rm eff}$), and a coverage of 0.05\,--\,0.4. Six more have hot spots, with temperatures up to 3000\,K above $T_{\rm eff}$ and coverage below 0.15. The remaining four stars have ambiguous solutions or are AA Tau-type contaminants. All of the stars with large spots (i.e. high coverage $>0.1$) are relatively cool with $T_{\rm eff} < 4500$\,K, which could be a result of having deeper convection zones. Apart from that, spot properties show no significant trends with rotation period, infrared excess, or stellar properties. Most notably, we find hot spots in stars that do not show $K-W2$ infrared excess, indicating the possibility of accretion across an inner disk cavity or the presence of plage.
\end{abstract}




\section{Introduction}

Variability is a key feature in young stellar objects (YSOs). One main reason why YSOs are more variable than main-sequence stars is that they are hosting strong magnetic fields, causing cool spots in interaction with the stellar photosphere and hot spots in interaction with the accretion flow from the disk.  While hot spots are only expected to appear on accreting stars (classical T Tauri stars) that have either full or transition disks (Class II objects), cool spots can be found on stars with and without disks. The presence of these spots, combined with the fast rotation of young stars, then induces periodic photometric variability on timescales of days (see reviews by \citet{2007prpl.conf..297H} and \citet{2014prpl.conf..433B}).

Photometric light curves, taken quasi-simultaneously in multiple bands have been an important tool to study the properties of spots for over 30 years \citep[see e.g.][]{1993A&A...272..176B, 1995A&A...299...89B, 2001AJ....121.3160C}. While light curves do not provide resolved pictures of the stellar surface like Doppler Imaging or related techniques \citep[e.g.][]{2002AN....323..309S, 2003A&A...408.1103S, 2009MNRAS.399.1829S, 2009ARA&A..47..333D}, they can convey a measurement of the temperature contrast between spots and stellar photosphere (thus allowing to distinguish between hot and cool spots), as well as the coverage, the fraction of the surface covered by spots \citep{1992ASPC...34...39S, 1995A&A...299...89B, 2009MNRAS.398..873S, 2012MNRAS.419.1271S}. In addition, multi-filter light curves can be obtained for large numbers of objects, helping to establish statistical trends. 

Variability induced by cool spots is typically limited to small amplitudes of less or around 10\,\%, but hot spots can produce much larger variability. We typically expect that hot spots are thousands of degrees warmer than the photosphere and very small, with a coverage in the range of a few percent \citep{1998ApJ...492..743M, 1998apsf.book.....H}, but more recent work has thrown doubts on these assumptions \citep{2009MNRAS.398..873S, 2012MNRAS.419.1271S, 2016MNRAS.458.3118B}. The temperature contrast between star and spot, as well as the spot coverage, are important parameters to verify and check models of  magnetospheric accretion \citep{2007prpl.conf..297H, 2014prpl.conf..433B}. It is important to note that light curves are only sensitive to spots distributed asymmetrically with respect to the rotational axes, as symmetric spot distributions do not cause variability.

The aim of this work is to establish a method that reliably recovers the properties of spots on YSOs from multi-filter broadband photometry. The Hunting Outbursting Young Stars (HOYS) citizen science project has been collecting such data of young clusters since 2014. We use the peak-to-peak amplitudes in at least three filters for stars with rotational variability. In contrast to previous work, we will build up an extensive grid of modelled solutions using atmospheric template spectra which are then compared with observed amplitudes, combined with a full treatment of the photometric errors. In doing so, laying the ground work to identify spots on many young stars in multiple clusters, and tracking spot properties over time. In this paper we present this method and findings for a sample of YSOs in IC~5070, identified in \cite{2021MNRAS.506.5989F}. This is the first survey to identify spot properties in this region. We detail our data in Sect.\,\ref{s_data} and discuss the methodology in Sect.\,\ref{methodology}, including the details of the systematic and statistical uncertainties. Our results are presented and discussed in Sect.\,\ref{results}.

\section{Data and YSO sample selection} \label{s_data}

\subsection{HOYS data and photometry} \label{sec_data_phot}

All photometry used in this project has been taken as part of the HOYS citizen science project \citep{2018MNRAS.478.5091F}. In this project, a combination of amateur, university, and professional telescopes are used to monitor the brightness of stars in nearby young clusters and star forming regions. The aim is to measure the brightness of all stars in optical broad-band filters every 12 to 24 hours.

The photometry in all images is calibrated against a deep reference frame taken under photometric conditions \citep{2018MNRAS.478.5091F}. These are obtained in the $u, B, V, R_C,$ and $I_C$ filters. The off-sets from instrumental to apparent magnitudes have been determined from the Cambridge Photometric Calibration Server\footnote{\tt \href{http://gsaweb.ast.cam.ac.uk/followup}{http://gsaweb.ast.cam.ac.uk/followup}}. These convert the magnitudes into the Johnson $U,B,V$ and Cousins $R_C, I_C$ system, which we will refer to as $U, B, V, R,$ and $I$ throughout the paper for simplicity.

Many of the HOYS images are taken in slightly different filters, such as the once used in DSLR cameras by some amateurs. Using these filters introduces colour terms in the photometry. In \citet{2020MNRAS.493..184E} we developed a procedure to correct these. We identify all non-variable stars in each field and use their known colours and magnitudes to determine the colour terms for each image and correct the ensuing photometry off-sets. This procedure is applied to all data used in our analysis. 

\begin{figure}
\centering
\includegraphics[width=\columnwidth]{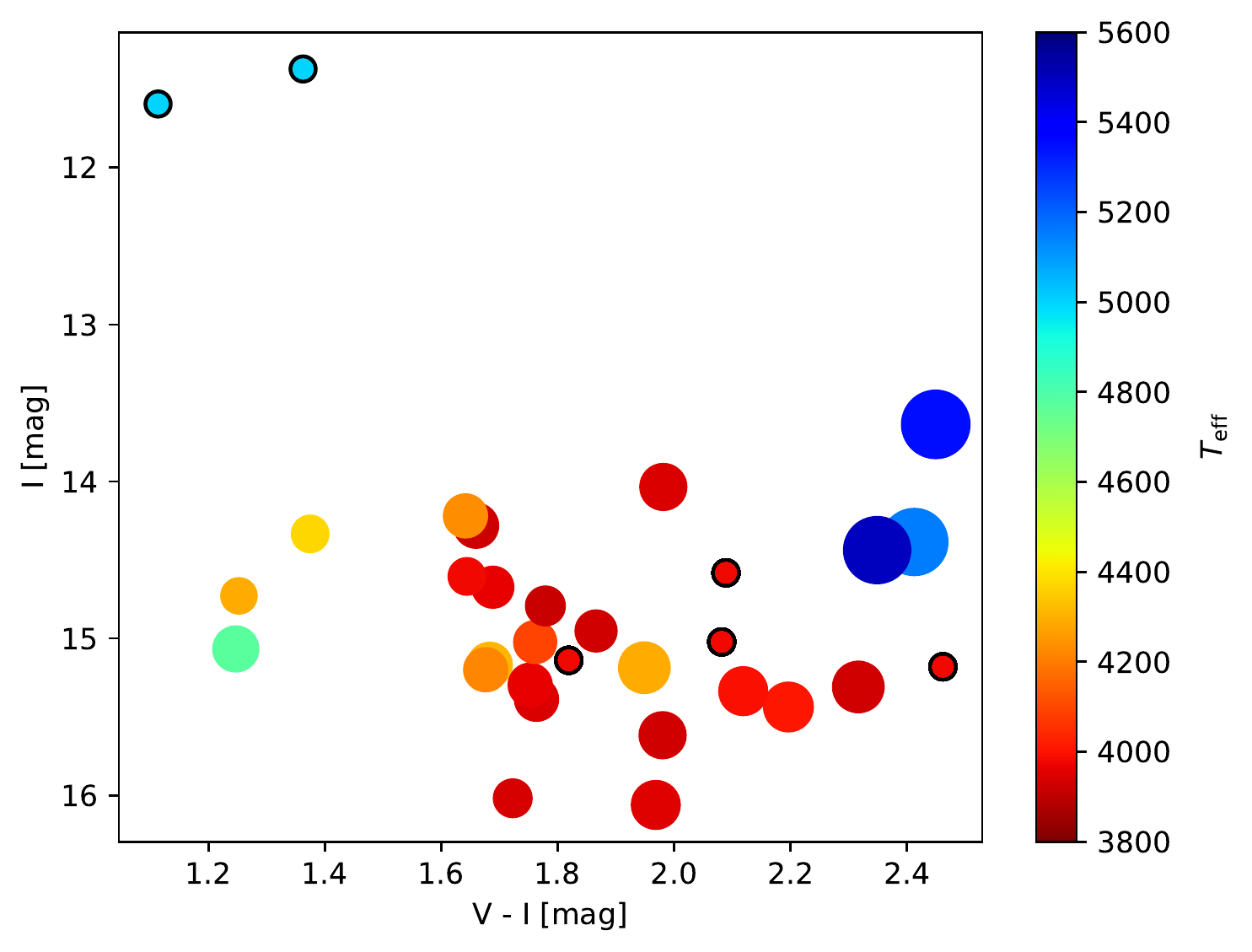}
\caption{ $V-I$ vs $I$ colour magnitude diagram of the YSO sample with data available in $V, R$, and $I$. We colour code the symbols with the effective temperature and the symbol size represents the optical extinction $A_V$ \citep{2020ApJ...904..146F}. For the objects without literature $T_{\rm eff}$ values we use our adopted values and indicated them by black rings around the symbols. No $A_V$ values are available for these. \label{cmdteff}}
\end{figure}

\subsection{YSO sample in IC~5070}

All targets investigated in this work are situated in the IC\,5070 star forming region (Pelican Nebula). Together with NGC\,7000 (North America Nebula) it is part of the large \hii\ region W\,80. The distance to the main cluster of YSOs in the region has been measured as 795\,pc by \citet{2020ApJ...899..128K}. The same authors identified a population of 395 young stars in the region belonging to six different dynamic groups. Most of the YSOs are about 1\,Myr old, and almost all of them have an age of less than 3\,Myr. 

\citet{2021MNRAS.506.5989F} investigated periodic variable objects in the IC\,5070 region. HOYS photometry taken over 80\,d in the summer of 2018 was utilised. The authors identified 59 periodic variables. Using data from GaiaEDR3, an unbiased sample of 40 YSOs in the nebula was identified, solely due to their periodicity, distance, and proper motion.  All phase folded light curves were presented in the Appendix of that work. \citet{2021MNRAS.506.5989F} found that the sample of periodic YSOs is split into fast and slow rotators, with a clear gap of periods around five days. Furthermore, the sample shows a 50/50 split of objects with and without disks, based on the $K-W2$ colours. Here we study the spot properties of these objects.

 For our analysis we require peak-to-peak measured amplitudes in at least three different filters,  $V$, $R$ and $I$ (see Sect.\,\ref{methodology}). This leaves 31 objects for investigation in the paper, which we define as our sample. Of those, 24 also include $B$ and 5 have all amplitudes from $U$ to $I$. Nine sources of the original sample have measurements only in $R$ and $I$. Many of the YSOs in this region have spectroscopic estimates of their effective temperatures ($T_{\rm eff}$) and spectral types in \citet{2020ApJ...904..146F}. Of the 31 YSOs in our sample, 25 have a counterpart in that work. 

We show a $V-I$ vs. $I$ colour magnitude diagram of all sources with $V, R$, and $I$ data available in Fig.\,\ref{cmdteff}. In this figure we colour code the symbols with the known $T_{\rm eff}$ from \citet{2020ApJ...904..146F} and the symbol size represents the optical extinction from the same work. One can see that the majority of low effective temperature (4000\,K) objects are spread out over a wide range of colours (1.6\,mag\,$\leq V-I \leq$\,2.5\,mag), due to varying extinction. Slightly higher temperatures (4400\,K) can be found for bluer colours. Interestingly, the three hottest stars are found at the reddest colours. These highly reddened sources are  most likely embedded YSOs or stars with their disk seen edge on. 

To estimate the effective temperature for the six objects without $T_{\rm eff}$ measurements in \citet{2020ApJ...904..146F}, we use the $V$-$I$ vs $I$ colour magnitude diagram shown in Fig.\,\ref{cmdteff}. Four of the six objects in question are at colours redder than 1.8\,mag, and for these we use the median $T_{\rm eff}$ value of 3979\,K of the other sources in this colour range. For the two brighter and bluer objects at about $I = 11.5$\,mag we adopt a higher value of 5000\,K as effective temperature, as this most closely resembles the values for objects with similar colours. We will discuss the influence of erroneous stellar temperatures on the determined spot properties in Sect.\,\ref{sec_kstars}. 

\subsection{Peak-to-peak amplitude determination} \label{sec_data_Tref}

For our analysis we require the measurement of the peak-to-peak amplitudes for the 31 periodic YSOs. This was done in the following way for all filters: we use the periods determined in \citet{2021MNRAS.506.5989F} to phase fold the photometry. A running median in phase space was determined, smoothed over 0.1 in phase. This smoothing includes on average 27 photometric data points from the typically 270 brightness measurements that are available in each filter per source. The peak-to-peak amplitude was determined as the difference between the minimum and maximum magnitude of the smoothed running median. We denote these as $\hat{A}_\lambda^o$, where $'o'$ indicates these refer to the observed values and $\lambda$ indicates the filter used.

The associated uncertainties $\sigma \left( \hat{A}_\lambda^o \right)$ are determined from the standard error of the mean of all photometric data points included in the determination of the minimum and maximum position of the running median. We list all investigated objects (ID numbers taken from \citealt{2021MNRAS.506.5989F}), their effective temperatures ($T_{\rm eff}$), periods, amplitudes, and uncertainties in Table\,\ref{tbl_ampe}. 

\begin{table*}
\caption{\label{tbl_ampe} Target list of all YSOs investigate in this work. For each object we list the ID numbers, the J2000 coordinates, their effective temperatures, the period, as well as the peak-to-peak amplitudes and uncertainties measured in the phase folded light curves. $^{(1)}$ from \citet{2021MNRAS.506.5989F}; $^{(2)}$ from \citet{2020ApJ...904..146F}. }
\centering
\setlength{\tabcolsep}{4pt}
\begin{tabular} {|c|cc|cc|cccccccccc|}
\hline
ID$^{(1)}$ & RA & DEC & $T_{\rm eff}$ $^{(2)}$ & Period$^{(1)}$ & $\hat{A}^o_{I}$ & $ \sigma \left( \hat{A}^o_{I} \right)$ & $\hat{A}^o_{R}$ &  $ \sigma \left( \hat{A}^o_{R} \right) $ & $\hat{A}^o_{V}$ & $ \sigma \left( \hat{A}^o_{V} \right)$ & $\hat{A}^o_{B}$ & $ \sigma \left( \hat{A}^o_{B} \right)$ & $\hat{A}^o_{U}$ & $ \sigma \left( \hat{A}^o_{U} \right)$ \\ 
 & [deg] & [deg] & [K] & [d] & [mag] & [mag] &  [mag] & [mag] & [mag] & [mag] & [mag] & [mag] & [mag] & [mag] \\ \hline
3220 & 313.37768 & 44.69840 & - & 0.866 & 0.052 & 0.012 & 0.057 & 0.017 & 0.089 & 0.022 & - & - & - & - \\
3988 & 312.72581 & 44.63562 & 3928 & 9.438 & 0.277 & 0.036 & 0.283 & 0.030 & 0.381 & 0.051 & 0.338 & 0.068 & - & - \\
4097 & 313.25278 & 44.61654 & 5350 & 1.683 & 0.061 & 0.012 & 0.077 & 0.019 & 0.075 & 0.037 & - & - & - & - \\
4446 & 313.10924 & 44.57396 & 5500 & 1.433 & 0.081 & 0.018 & 0.094 & 0.017 & 0.080 & 0.024 & 0.113 & 0.050 & - & - \\
4766 & 312.75374 & 44.53048 & 4091 & 6.602 & 0.667 & 0.046 & 0.606 & 0.032 & 0.817 & 0.041 & 1.066 & 0.080 & - & - \\
5535 & 312.83745 & 44.43877 & 3921 & 3.862 & 0.233 & 0.020 & 0.230 & 0.016 & 0.258 & 0.016 & 0.282 & 0.029 & 0.536 & 0.221 \\
5548 & 312.99661 & 44.42881 & 3943 & 4.157 & 0.047 & 0.012 & 0.063 & 0.012 & 0.071 & 0.016 & 0.075 & 0.026 & - & - \\
5559 & 312.93711 & 44.43862 & 3940 & 3.759 & 0.157 & 0.030 & 0.201 & 0.026 & 0.213 & 0.034 & 0.282 & 0.060 & - & - \\
5575 & 313.41700 & 44.43060 & 5150 & 1.390 & 0.053 & 0.015 & 0.086 & 0.028 & 0.104 & 0.047 & - & - & - & - \\
5886 & 312.12005 & 44.40321 & - & 9.041 & 0.080 & 0.025 & 0.137 & 0.047 & 0.159 & 0.082 & - & - & - & - \\
6060 & 312.81885 & 44.38279 & 4291 & 2.427 & 0.075 & 0.018 & 0.104 & 0.012 & 0.116 & 0.012 & 0.130 & 0.016 & 0.181 & 0.073 \\
6149 & 312.94395 & 44.37257 & 3928 & 2.176 & 0.179 & 0.023 & 0.222 & 0.022 & 0.257 & 0.029 & 0.306 & 0.098 & - & - \\
6259 & 312.77765 & 44.36132 & 4775 & 1.398 & 0.095 & 0.020 & 0.107 & 0.014 & 0.112 & 0.014 & 0.119 & 0.021 & - & - \\
6315 & 313.07439 & 44.35443 & 3952 & 3.223 & 0.114 & 0.038 & 0.141 & 0.030 & 0.178 & 0.079 & - & - & - & - \\
6337 & 312.84446 & 44.35212 & 3964 & 3.911 & 0.313 & 0.028 & 0.290 & 0.024 & 0.296 & 0.025 & 0.308 & 0.054 & - & - \\
6393 & 313.35269 & 44.34279 & 3990 & 2.773 & 0.158 & 0.029 & 0.194 & 0.047 & 0.277 & 0.074 & - & - & - & - \\
6813 & 312.81307 & 44.30490 & 3946 & 4.167 & 0.091 & 0.022 & 0.143 & 0.021 & 0.165 & 0.022 & 0.250 & 0.054 & - & - \\
6861 & 313.04822 & 44.29854 & 4292 & 3.522 & 0.111 & 0.024 & 0.132 & 0.019 & 0.139 & 0.025 & 0.144 & 0.044 & - & - \\
6929 & 312.74460 & 44.29190 & 3916 & 7.276 & 0.122 & 0.017 & 0.144 & 0.016 & 0.166 & 0.017 & 0.177 & 0.040 & - & - \\
7181 & 312.75654 & 44.26168 & 3979 & 7.338 & 0.158 & 0.018 & 0.175 & 0.015 & 0.196 & 0.016 & 0.174 & 0.027 & - & - \\
7422 & 312.74312 & 44.24232 & 4373 & 4.901 & 0.058 & 0.017 & 0.059 & 0.010 & 0.066 & 0.011 & 0.082 & 0.015 & 0.144 & 0.079 \\
7465 & 313.14529 & 44.23348 & 4216 & 10.573 & 0.126 & 0.024 & 0.162 & 0.019 & 0.226 & 0.026 & 0.231 & 0.054 & - & - \\
7472 & 313.09386 & 44.23339 & 4311 & 3.049 & 0.120 & 0.026 & 0.147 & 0.019 & 0.181 & 0.026 & 0.199 & 0.043 & - & - \\
7632 & 312.82600 & 44.21895 & 3966 & 7.853 & 0.229 & 0.019 & 0.281 & 0.014 & 0.312 & 0.016 & 0.331 & 0.027 & - & - \\
7954 & 313.35736 & 44.17926 & 4010 & 1.449 & 0.103 & 0.030 & 0.128 & 0.047 & 0.194 & 0.067 & - & - & - & - \\
8025 & 312.45491 & 44.17952 & - & 3.313 & 0.342 & 0.022 & 0.444 & 0.019 & 0.484 & 0.024 & 0.575 & 0.047 & - & - \\
8038 & 312.78141 & 44.17628 & - & 3.522 & 0.095 & 0.020 & 0.118 & 0.017 & 0.140 & 0.024 & 0.157 & 0.047 & - & - \\
8249 & 312.76358 & 44.15360 & 3928 & 7.880 & 0.067 & 0.019 & 0.092 & 0.017 & 0.103 & 0.023 & 0.076 & 0.036 & - & - \\
9267 & 312.87064 & 44.07309 & - & 4.830 & 0.133 & 0.010 & 0.137 & 0.009 & 0.141 & 0.006 & 0.157 & 0.006 & 0.132 & 0.018 \\
9321 & 312.87737 & 44.06251 & 4235 & 3.166 & 0.174 & 0.017 & 0.214 & 0.014 & 0.248 & 0.016 & 0.304 & 0.021 & 0.266 & 0.083 \\
9961 & 313.09561 & 44.01582 & - & 3.625 & 0.053 & 0.019 & 0.062 & 0.017 & 0.050 & 0.023 & 0.073 & 0.046 & - & - \\
\hline
\end{tabular}
\end{table*}

\section{Spot fitting Methodology}\label{methodology}

In order to recover the spot properties from the observed peak-to-peak amplitudes of our YSO sample, we have modelled amplitudes by generating star-spot systems with synthetic spectra. 

\subsection{Basics}

For the analysis of the periodic variable objects in our sample, we assume that the young stars have one dominant spot, i.e. they represent the most simple system of star and spot. Hence, we assume a star with an effective temperature $T_\star$, no limb darkening, and a uniform spot with a temperature $T_S$, which covers a fraction $f$ of the visible stellar surface. Thus, we only investigate the peak-to-peak brightness variations.  This most basic of approaches goes back to e.g. \citet{1995A&A...299...89B} and \citet{2001AJ....121.3160C}. However, here we apply it simultaneously to a multi-filter data set, use atmospheric models as opposed to assuming black body emission, and accurately determine the reliability and accuracy of this method.  

We acknowledge that we are operating with a very simplified model in assuming a uniform spot temperature and disregarding limb darkening. Limb darkening causes a systematic reduction in the stellar flux, that would lead to an underestimation in spot size in the order of a few percent. As we will show, this is significantly smaller than our statistical uncertainties. By considering a single uniform spot, we are removing some complexity and focusing on the dominant properties of the asymmetric component of the spot distribution. It is beyond the scope of this paper to model the shape of the phase-folded light curves to infer the spot latitude on the star and inclination of the stellar rotation axis. This would be influenced by limb darkening.

The ratio of the flux from the un-spotted surface in a filter $\lambda$, and the spotted surface corresponds to the peak-to-peak amplitude of the variation. The geometry of the system is simplified as either considering the spot to be visible when `in front' or not visible when `behind'. Thus, the model peak-to-peak amplitude $\hat{A}^m_\lambda$ in each filter can be determined by Eq.\,\ref{eq_A_spotstar_c}. 

\begin{equation}
    \hat{A}^m _{\lambda}  =  \left| 2.5 \cdot \log \left( \frac{F^\lambda \left( T_\star \right)}{\left( 1 - f\right) \cdot F^\lambda \left(T_\star \right) + f \cdot F^\lambda (T_S) } \right) \right|
    \label{eq_A_spotstar_c}
\end{equation}

Two sets of synthetic stellar atmosphere spectra (PHOENIX and ATLAS9) have been used to determine the fluxes $F^\lambda(T)$. They have been accessed through the astropy {\tt PySynphot} distribution \citep{2013ascl.soft03023S}. To obtain the fluxes for each filter, we have used the {\tt speclite.filters}\footnote{\tt \url{https://github.com/desihub/speclite/blob/master/speclite/filters.py}} package, and convolved the atmospheric model spectra with transmission curves for $U, B, V, R,$ and $I$. 

The PHOENIX models \citep{2013A&A...553A...6H} are a library of synthetic spectra for effective temperatures from  2000~K to 70000~K, metallicities [M/H] from $-4.0$ to $+5.0$ and surface gravities $\log(g)$ from 0.0 to 6.0. The {\tt PySynphot} distribution uses models obtained in 2011, although in 2021 the grid was updated to models from \citet{2013A&A...553A...6H} for [M/H]~=~0.0, with updates for the other metallicities planned. This update has an improved resolution, which makes it more suited for spectroscopic work. In this work they are convolved with broadband filters. 

The ATLAS9 grid \citep{2003IAUS..210P.A20C} is an update to previous models published in \citet{1991ASIC..341..441K}. The grid covers effective temperatures from 3500~K to 50000~K, metallicities from $-2.5$ to $+0.5$, and $\log(g)$ from 0.0 to 5.0. The lower temperature limit is above that of the PHOENIX models, and limits the application to cold spots on our YSOs as there is less parameter space below the stellar temperature. 

The PHOENIX models and the ATLAS9 grid are compared in \citet{2013A&A...553A...6H}. The relevant points of interest for this work are that the PHOENIX models were generated using solar abundances from \citet{2009ARA&A..47..481A}, whereas the ATLAS9 grid used solar abundances from \citet{1998SSRv...85..161G}. The two models take different approaches to convection, which affects the line profiles. PHOENIX uses a mixing length parameter $(\alpha)$ for macroturbulence varied according to temperature, following \citet{1958ZA.....46..108B}. The low temperatures used in this work have a mixing length parameter of $\alpha$ between 1.5 and 3. The ATLAS9 models use a constant value of $\alpha$~=~1.5, and adjust for non-convective overshooting below 7000~K. Pre-main sequence stars are fully convective until they reach sufficient mass to generate a radiative core. 

As we will see below, the ATLAS9 grid is less well suited for our purpose than the PHOENIX models due to the 3500~K lower effective temperature limit. This does not enable the accurate determination of cold spot properties on low effective temperature objects. Thus, unless stated otherwise, all analysis, discussion, and results presented in the paper are obtained using the PHOENIX model stellar atmospheres. We have, however, performed the entire analysis for both sets of models. The results for both model sets agree for the high effective temperature objects, showing the results do not depend on the stellar atmosphere model used. 

\subsection{A Note on Notation \label{sec_anoteonnotation}}

Our data of the YSOs, i.e. the measured peak-to-peak amplitudes, have been obtained in multiple filters. In the analysis we compare these measurements with the model calculations for different sub-samples (sets) of amplitudes. These amplitude sets hence refer to the same model or observation in different filters. Although any combination of filters can be used, as discussed in Sect.\,\ref{sec_data_phot}, the filters are used strictly sequentially from short to long wavelengths. The minimum data requirement for inclusion in the dataset are peak-to-peak amplitudes $\hat{A}_V$, $\hat{A}_R$, and $\hat{A}_I$. The notation we will use hereafter for such a set is $\hat{A}_{\{V\}}$, where $V$ denotes the shortest wavelength filter included in a dataset. In this way $ \hat{A}_{\{U \}} = \{ \hat{A}_U, \hat{A}_B, \hat{A}_V, \hat{A}_{R}, \hat{A}_{I}  \}$. A superscript when present refers to the origin of the amplitude, such as modelled $\hat{A}^m_{\{ \lambda \}}$ in Eq. \ref{eq_A_spotstar_c}. 

\subsection{Identification of best Spot Model}\label{bestmodel}

To match observed amplitudes to model spot properties, a large number of peak-to-peak amplitude sets were modelled following Eq.\,\ref{eq_A_spotstar_c}. The YSO effective temperature range of our sample (as discussed in Sect.\,\ref{sec_data_Tref}) is 3800~K~--~5500~K. Equation~\ref{eq_A_spotstar_c} requires a fixed stellar temperature to create a set of model amplitudes. We created these model amplitude sets for effective stellar temperatures in steps of 50~K intervals. This saves significant computing time as the same amplitude sets can be used for several objects. For each object we choose the model set closest in effective temperature and show in Sect.\,\ref{sec_kstars} that this has no significant influence on the results. 

The investigated parameter space for the spot temperature is $2000~\rm{K} \leq T_S \leq 12000~\rm{K}$ (PHOENIX) and $3500~\rm{K} \leq T_S \leq 12000~\rm{K}$ (ATLAS9). For the spot coverage we investigate $0 \leq f \leq 0.5$. These ranges were homogeneously, randomly sampled $10^6$ times for each stellar temperature. This corresponds to an average spacing in spot temperature of 10~K and 0.0005 in spot coverage.

Solar metallicity and $\log(g)$ = 4.0 were assumed throughout, which are reasonable values for our sample. The systematic uncertainties when altering these parameters are less than the statistical uncertainties of our procedure. We describe this in more detail in Appendix~\ref{sec_systematics}.

We use Eq.\,\ref{eq_rms_def} to determine the separation between the observed amplitude $\hat{A}_{\{\lambda\}} ^o$ and the modelled amplitude set $\hat{A}_{\{\lambda\}} ^m$. There, $N$ is the length of the filter set $\{ \lambda \}$, i.e. $N = 3$ for $\{V\}$. The observed amplitudes $ \hat{A}^o_{ \{\lambda \}}$ are compared to each of the $10^6$ modelled amplitude sets $\hat{A}^m_{\{ \lambda \}}$. The model with the minimum $RMS_{ \{ \lambda \} }$ is the best fitting spot model for the observations. 

\begin{equation}
    RMS_ {\{\lambda \} }=  \sqrt{ \frac{1}{N}\sum_{ \{\lambda \}} \left( \hat{A}^o_{\lambda}  - \hat{A}^m_{\lambda} \right)^2}
    \label{eq_rms_def}
\end{equation}

\begin{figure*}
\centering
\includegraphics[width=1.01\columnwidth]{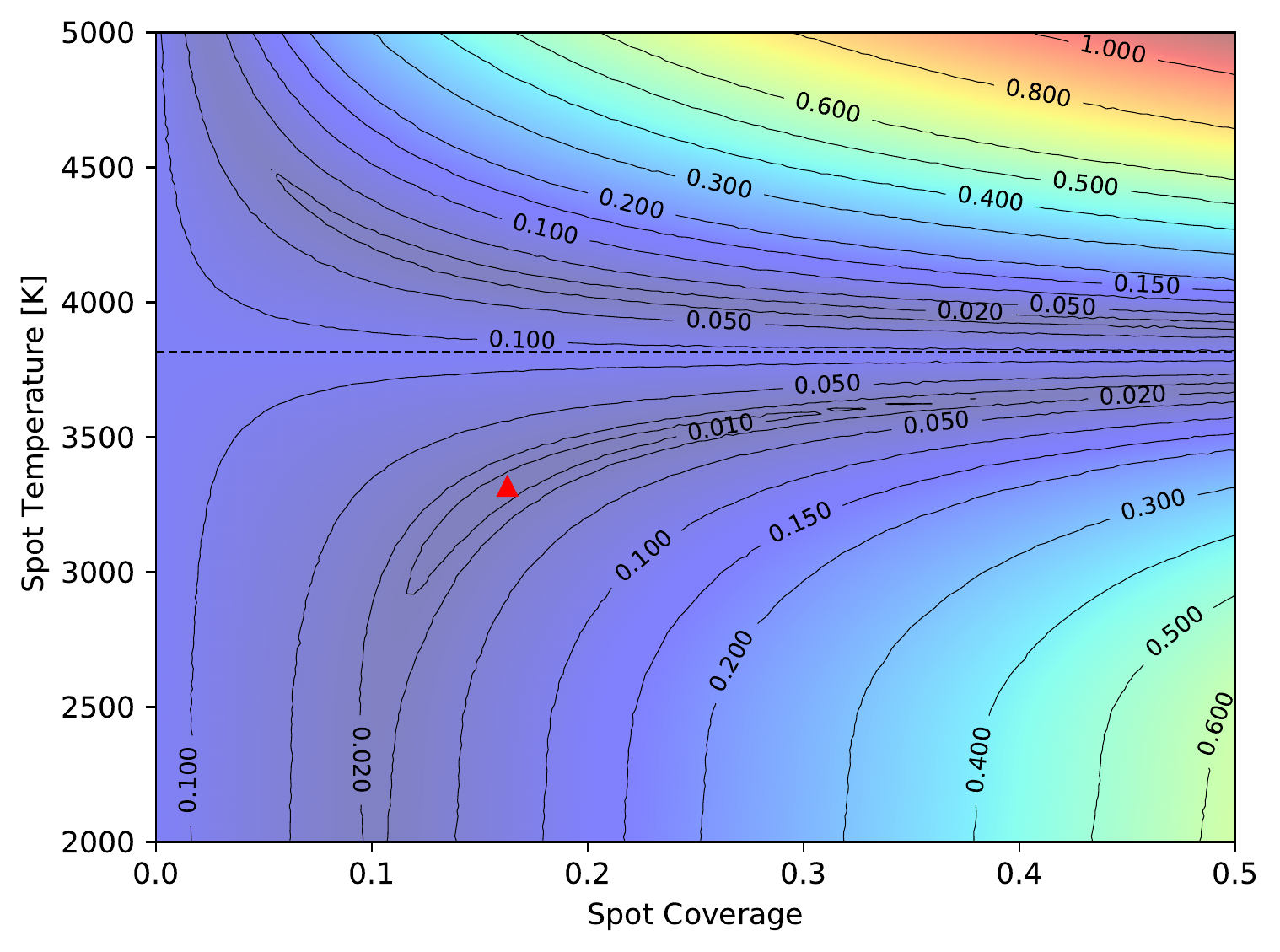} \hfill
\includegraphics[width=0.97\columnwidth]{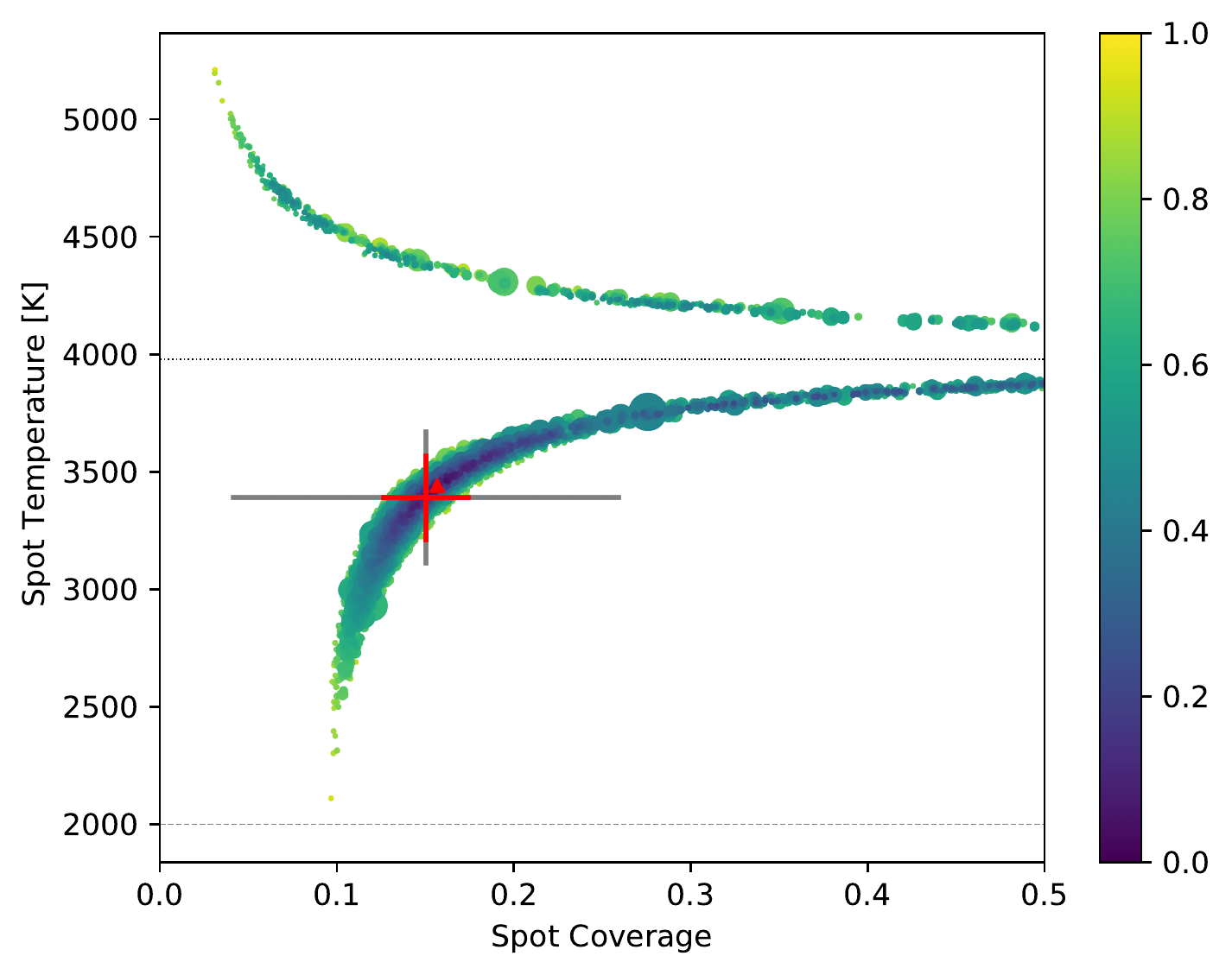} \hfill
\caption{  {\bf Left:} Contour plot showing $RMS_{\{V\}}$ between the observed amplitudes  $\hat{A}^o_{ \{ V \}}$ and the modelled amplitudes $\hat{A}^m_{ \{ V \}}$ for object 8038. A red triangle indicates the model with the absolute minimum  $RMS_{\{V\}}$, i.e. the best fitting model. {\bf Right:} The best fitting models for 10000 sets of amplitudes $\hat{A}^v_{ \{ V \}}$, for object 8038 varied within their associated uncertainties. The colour bar is the measure of the variation $RMS^{\sigma}_{\{\lambda\}}$ and the symbol size is inversely proportional to the minimum RMS. In this case $87.78$\,\% of the recovered models are cold spots. Positioned on the median temperature and spot values, the grey cross is the standard deviation of cold spot solutions, and the red cross is the median absolute deviation for the cold spots. The red triangle is the best fitting model from the observed amplitudes. The dotted line indicates the stellar temperature, and the dashed line the minimum temperature available in the atmospheric models. \label{fig_8038}}
\end{figure*}

In the left panel of Fig.~\ref{fig_8038} we show a contour map of $RMS_{ \{ V \} }$ for object 8038 as designated in \citet{2021MNRAS.506.5989F} for the PHOENIX models. The absolute minimum in $RMS_{ \{ V\} }$ is marked by a red triangle. The contour map shows two local minima in the regions above and below the stellar temperature. The lowest contours on both, the cold and warm spot solution, roughly follow reciprocal lines between spot temperature and spot coverage, i.e. $(T_S-T_\star) \propto \pm 1/f$. The minimum in the cold spot region has a value around $RMS_{\{V\}}  = 0.01$~mag, whereas the minimum value in the hot spot region is $RMS_{\{V\}} = 0.02$~mag. For the majority of our objects, one minimum (hot or cold spot) is significantly deeper than the other. 

Values near the minimum $RMS_{\{\lambda\}}$ can cover a large area of the parameter space. In the case of Fig.~\ref{fig_8038}, the $RMS_{\{V\}}  = 0.01$~mag contour covers a range of about 500~K in spot temperature and 0.2 in coverage. To estimate the uncertainties of our solution for the spot properties as well as to establish if the hot or cold spot solution dominates, the statistical uncertainties of the measured amplitudes need to be taken into account. 

\subsection{Statistical uncertainties of best spot model \label{sec_statistical}}

The position of the absolute minimum of $RMS_{\{V\}}$ can be sensitive to small perturbations in the measured amplitudes, and is affected directly by the resolution of the model amplitude set. To consider the statistical uncertainties in the methodology the best model fitting was applied to amplitudes that are varied within their measurement errors $\sigma \left( \hat{A}_\lambda^o \right)$, established in Sect.~\ref{sec_data_Tref}. 

The observed amplitudes $\hat{A}^o_\lambda$ for each object were varied with a Gaussian distribution according to their associated uncertainties. This generates sets of varied amplitudes $\hat{A}^v_{\{\lambda \}}$. We reduce these variations to those that are within $1\sigma$ of the measured amplitude in each filter. Our tests have shown that the values of the spot properties and their uncertainties stabilise at 3000 iterations or more. Thus, we have created 10000 iterations of $\hat{A}^v_{\{\lambda \}}$ within their $1\sigma$ uncertainties for all objects. 

As the amplitude in each filter was varied independently, to determine the extent of the perturbation upon $\hat{A}^o_{\{ \lambda \}}$, a measure $RMS^{\sigma}_{\{\lambda\}}$ is defined in Eq.~\ref{eq_rms_sig}. A low value for $RMS^{\sigma}_{\{\lambda\}}$ indicates little variation. In cases where $RMS^{\sigma}_{\{\lambda\}}$ is close to one, each of the amplitudes in the set has been varied close to $1\sigma$. 

\begin{equation}
    RMS^{\sigma}_{\{\lambda\}} = \sqrt{\frac{1}{N}\sum_\lambda \frac{\left( \hat A_\lambda ^o - \hat A_\lambda^v \right)^2}{\sigma^2 \left( \hat{A}_\lambda^o \right)} }
    \label{eq_rms_sig}
\end{equation}

The best fitting spot properties for each $\hat{A}^v_{\{\lambda \}}$ were determined by identifying the minimum $RMS_{\{\lambda\}}$ between $\hat{A}^v_{\{\lambda \}}$ and $\hat{A}^m_{\{\lambda \}}$. The distribution of best fit spot properties from the 10000 sets of $\hat{A}^v_{\{\lambda \}}$ shows a clear trend towards cold or hot spots for the majority of targets. In a small number of cases, the best fit spots do not show a clear preference to hot or cold. To indicate cases where the solution is ambiguous, we determine the ratio of hot to cold spot solutions HS:CS$_{\{\lambda\}}$. If the value is close to zero or one, the object is clearly a cold or hot spot, respectively. In cases where the ratio is close to 0.5, the solution is ambiguous. The ratio depends on the filter set used and the signal to noise ratio of the measured amplitudes. We list the HS:CS$_{\{\lambda\}}$ values and the associated signal to noise values of the amplitudes in Table\,\ref{tbl_keynumbers}. They are discussed in more detail in Sect.\,\ref{sec_IC5070targets}. For the purpose of this work, objects with $0.4 < $~HS:CS$_{\{\lambda \}} < 0.6$ are considered ambiguous, and are removed from any statistical analysis. These objects are identified in Table\,\ref{tbl_keynumbers}. There we also list all HS:CS$_{\{V\}}$,  HS:CS$_{\{B\}}$, and HS:CS$_{\{U\}}$ values for available data. We will discuss the influence of the choice of threshold on the statistics in our sample in Sect.\,\ref{sec_IC5070targets}.

The best fitting spot properties and their uncertainties are determined using only the solutions (hot or cold spot) where the majority of the best fits are situated. The median and median absolute deviation (MAD) of those were determined for the spot temperature and spot coverage, and their respective errors. Figure~\ref{fig_8038} (right) shows the distribution of best models recovered for the $10000$ iterations of $\hat{A}^v _ {\{V\}}$ for object 8038. The object has HS:CS$_{\{V\}} = 0.1222$ (see Table \ref{tbl_keynumbers}). The results from the PHOENIX models are shown. When repeated with the ATLAS9 grid, the best models cluster around the lower temperature limit of the model at 3500~K. The model with the lowest $RMS_{\{V\}}$ from $\hat{A}^o_{\{V \}}$ is shown again as a red triangle. Figure\,\ref{fig_8038} (right) also shows the standard deviation as an grey cross for comparison and the MAD as a red cross. We adopt the MAD as the uncertainty measurement instead of the standard deviation because the spot temperature and spot coverage are correlated. Hence, the standard deviation is not a good representation of the uncertainty. The MAD is always significantly smaller than the standard deviation, especially in spot coverage. The measure of dispersion $RMS^{\sigma}_{\{\lambda\}}$, indicated by the colour bar in the figure, shows that the $\hat{A}^v_{ \{ V \}}$ values that have been varied less are grouped around the median. In this example, the hot spot solutions are produced by $\hat{A}^v_{ \{ V \}}$ with high $RMS^{\sigma}_{\{V\}}$.

\subsection{Systematic uncertainties of spot properties \label{sec_kstars}}

In the previous sections we have established our methodology to infer spot properties and their statistical uncertainties from a set of peak-to-peak amplitudes. In this section we investigate potential systematic uncertainties. There are three main sources for these: i) the choice of stellar properties ($T_{\rm eff}$, $\log(g)$, [M/H]); ii) the correlation of spot temperature and coverage; iii) the choice of filters included in the analysis. We discuss the details of the evaluation of the systematic uncertainties in Appendix\,\ref{sec_systematics} and summarise the main results here. 

i) Changes in the assumed stellar properties in the models do result in small systematic changes in the determined spot properties. In almost all cases these shifts are significantly smaller than the statistical uncertainties determined in the previous section. In particular changing the metallicity and surface gravity within one dex, or the effective temperature within 500\,K, causes shifts in the determined properties that are smaller than the statistical uncertainties.

ii) Using our methodology we can determine amplitudes for spots with simulated properties. We have in turn used these amplitudes to recover the properties of these simulated spots. The resulting spot properties are systematically shifted, due to the correlation of spot temperature and coverage. For cold spots the recovered properties result in slightly smaller spots which are colder than the model spot temperature. For hot spots the size is also slightly reduced and the spot temperatures are increased compared to the input values. In all cases, these systematic off-sets are below the associated statistical uncertainties.

iii) Including amplitudes at shorter wavelengths ($B$ and $U$) does decrease the systematic shifts for all spots. It also reduces the statistical uncertainties for hot spots. But for cold spots the statistical uncertainties increase. This is due to the typical temperatures of the stars and spots in our sample. Thus, since most of our sample shows cold spots (see Sect.\,\ref{sec_IC5070targets} below), it is not a disadvantage that for the majority of our sources we do not have $B$ or $U$ amplitudes available for analysis.

\begin{figure*}
\centering
\includegraphics[width=\columnwidth]{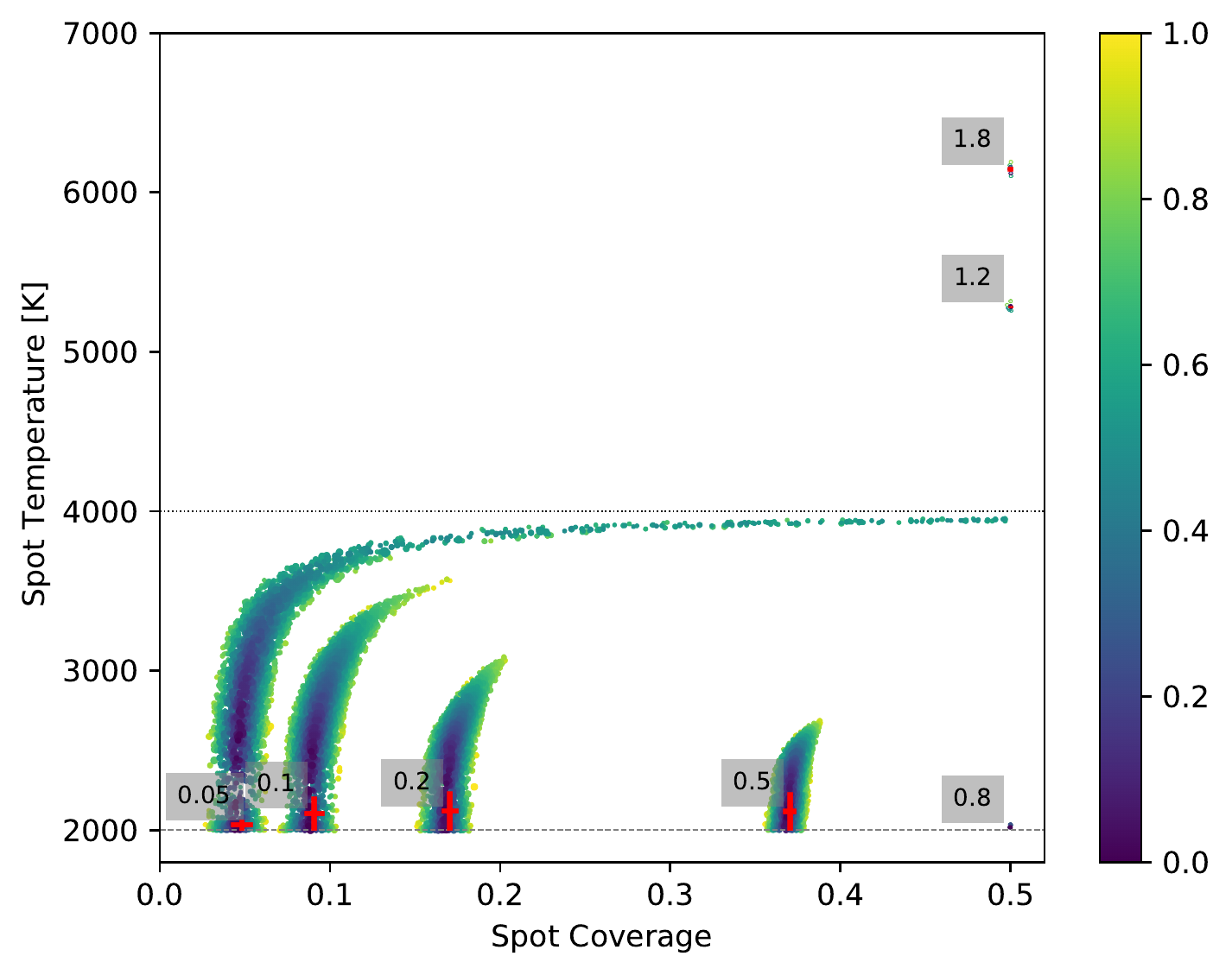} \hfill
\includegraphics[width=\columnwidth]{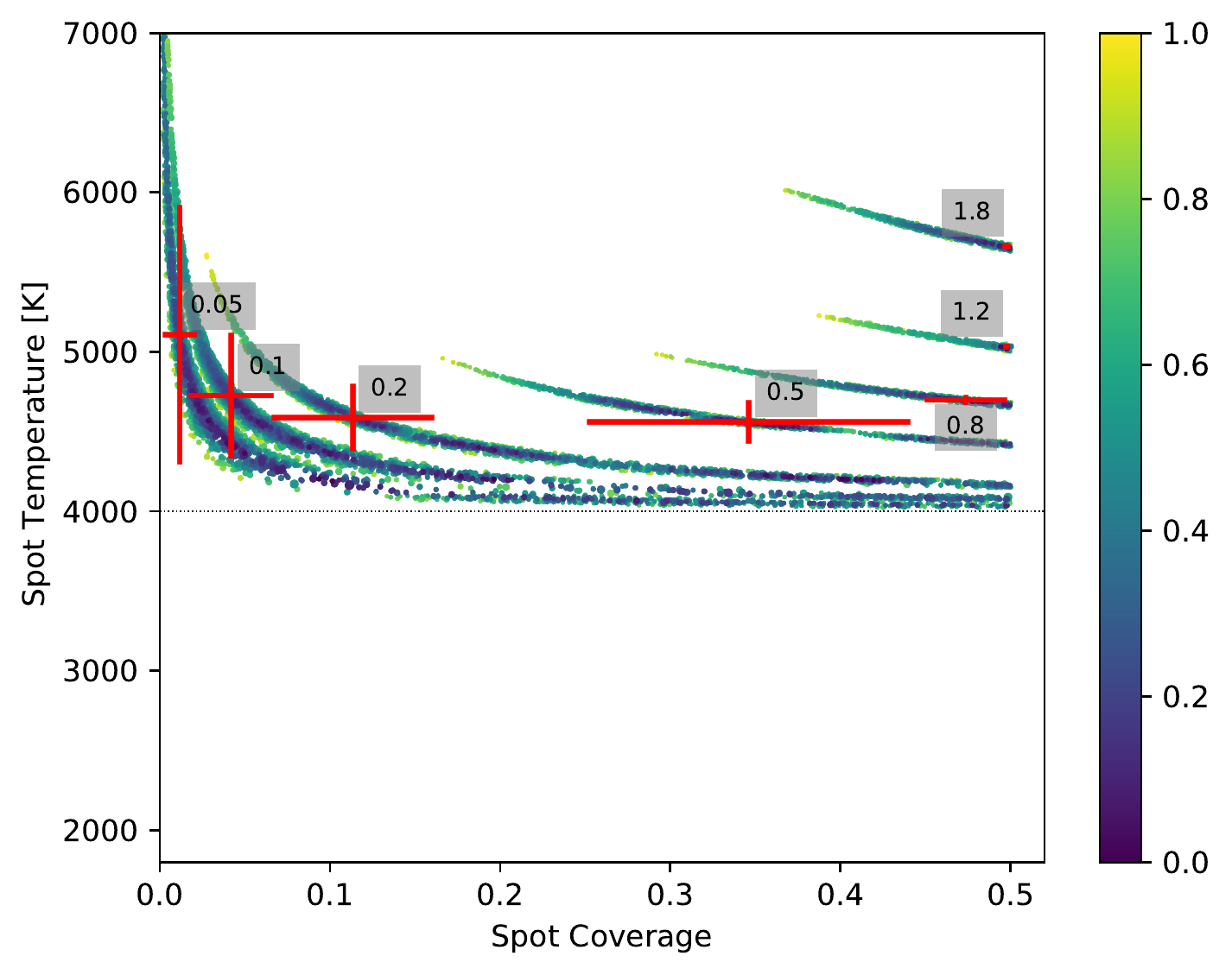}
\caption{Spot properties estimated for simulated grey (left) and $R_V = 5.0$ (right) extinction caused by AA~Tau like sources. The red crosses mark the position and uncertainties of the spot properties for $\hat{A}^v _{\{V \}}$. The values in the grey boxes indicate the $\hat{A}_V$ amplitudes. The coloured symbols follow the same scheme as in the right panel of Fig.\,\ref{fig_8038}. \label{extinct_V} }
\end{figure*}

\subsection{Identification of Non-Spot contaminants}

Periodic variability with light curves of similar appearance to spot modulation, can also be caused by other sources. The nature and selection of our sample excludes non-YSO causes such as pulsations in giant stars. However, the periodic occultation of the central star by inner disk material can mimic a light curve of a spotted star. The prototypical object for this kind of variation is AA\,Tau \citep{1999A&A...349..619B}. This is a T\,Tauri star which undergoes quasi-periodic dimming of $\approx 1.4$\,mag in $B,V,R,$ and $I$ due to a warp in the inner disk. To identify variability caused by such occultations in our YSO sample, we have applied the same spot recovery method as the one we used to determine the systematic uncertainties with simulated spots in Sect.\,\ref{sec_kstars}. Here we are using amplitudes artificially generated according to various extinction laws.

Three models of extinction have been tested: $R_V = 3.1$ and $R_V = 5.0$ modelled from \citet{1990ARA&A..28...37M}, and grey extinction. The latter indicates equal extinction in all filters caused by large ($>> 1 \mu m$) dust grains or opaque material. Amplitudes in the different filters have been generated according to the three extinction models and the spot properties and uncertainties were calculated using the same method as laid out in Sect.\,\ref{sec_statistical} using $ A^m_{\{V\}}$, the PHOENIX models, a stellar temperature of $4000~\rm{K}$, and the average associated errors for our YSO sample (0.02\,--\,0.03\,mag, depending on the filter). We show the recovered spot properties for grey extinction and $R_V = 5.0$ in Fig.\,\ref{extinct_V}. In all cases we use $\hat{A}_{\{V \}}$ and the amplitudes in the visual filter are $A_V$~=~0.05, 0.1, 0.2, 0.5, 0.8, 1.2, and 1.8~mag. For clarity in Fig.\,\ref{extinct_V}, we only show the data points that correspond to the solution (hot/cold spot) where the majority of the 10000 iterations fall. For the gray extinction model, the HS:CS$_{\{V \}}$ is basically zero or one in all cases. The HS:CS$_{\{V \}}$ for the $R_V = 5.0$ simulation is close to 0.6 for $V$-band amplitudes below 0.2\,mag and one for the others.

From Fig.\,\ref{extinct_V} we see that for grey extinction the inferred spot temperatures cluster near the minimum temperature available, as long as the amplitudes are small (less than half a magnitude). For larger amplitudes, the properties cluster near the upper limit of the spot coverage investigated. The former is understandable as the grey extinction essentially blocks the same small fraction of light in all filters. This is equivalent to a small, completely black ($T$\,=\,0\,K) object transiting the star. The $R_V = 5.0$ reddening leads to solutions that are mimicking hot spots. The coverage increases with the amplitude in $V$ and the spot temperature increases with $\hat{A}_V$ when the coverage reaches the limit of the investigated parameter space. With the exception of very small amplitudes, the inferred spot coverage is much larger than what can be expected for hot spots. Furthermore, the amplitudes for hot spots with the same temperature and coverage are much higher than for extinction changes.

The results for $R_V = 3.1$ (not shown) are similar to the $R_V = 5.0$ case, with slightly smaller inferred coverage values. Note that changing the stellar properties or using the ATLAS9 models leads to the same results. Thus, our method is able to identify non-spot contaminants in our sample. They are selected as either showing up as very cold spots (at the minimum temperature investigated), unrealistically large spots (coverage near 0.5), or hot spots with very large coverage and/or amplitudes too small to be associated with a hot spot. Based on this, we have identified and removed two objects from our sample as a likely AA~Tau objects. The properties of these are consistent with having grey extinction. They are indicated in Table\,\ref{tbl_keynumbers} and discussed in Sect.\,\ref{sec_IC5070targets}. 

\begin{table*}
\caption{\label{tbl_keynumbers} The recovered spot properties and uncertainties for our YSO sample from Table \ref{tbl_ampe}. The objects are listed based on our adopted classification as hot/cold spot, ambiguous source, or potential AA~Tau contaminant. We list the ID number, the used effective temperature of the star during the modelling, the spot properties (temperature and coverage), and their MAD uncertainties obtained from the $\{V\}$ data. In the last six columns we list the HS:CS ratios and the minimum signal to noise ratio of the associated amplitudes (see text for details). Objects marked with a (*) have no effective temperature reference in \citet{2020ApJ...904..146F}, and we have adopted the sample mean (rounded to 4000\,K) or 5000\,K for the two bright objects. $^{(1)}$ from \citet{2021MNRAS.506.5989F}; $^{(2)}$ from \citet{2020ApJ...904..146F}.}
\centering
\begin{tabular} {|cc|cccc|cccccc|}
\hline
ID$^{(1)}$ & $T_{\rm eff}$ $^{(2)}$ & $T_S $ & $T_S$ MAD & $f$ & $f$ MAD & HS:CS$_{\{V\}}$ & Min SNR$_{ \{V \}}$  & HS:CS$_{\{ B\}}$& SNR$_{B}$ &HS:CS$_{\{ U\}}$ & SNR$_{U}$ \\ 
 & [K] & [K]& [K]& & & & & & & & \\ \hline
\multicolumn{12}{|l|}{Hot spot in HS:CS$_{\{ V \}}$} \\ \hline
3220* & 5000 & 6643 & 753 & 0.026 & 0.014 & 0.8168 & 3.43 &  -  &  - &  -  &  - \\
5575 & 5150 & 7798 & 972 & 0.021 & 0.010 & 0.6829 & 2.21 &  -  &  - &  -  &  - \\
5886* & 4000 & 4910 & 609 & 0.054 & 0.042 & 0.6024 & 1.95 &  -  &  - &  -  &  - \\
6393 & 4000 & 5182 & 384 & 0.066 & 0.029 & 0.6029 & 3.71 &  -  &  - &  -  &  - \\
7465 & 4200 & 4882 & 184 & 0.125 & 0.049 & 0.8879 & 5.29 & 0.4081 & 4.28 &  -  &  - \\
7954 & 4000 & 5372 & 579 & 0.032 & 0.018 & 0.6787 & 2.72 &  -  &  - &  -  &  - \\ 
\hline
\multicolumn{12}{|l|}{Cold spot in HS:CS$_{\{ V \}}$} \\ \hline
3988 & 3950 & 3139 & 112 & 0.311 & 0.025 & 0.0557 & 7.49 & 0.0011 & 4.97 &  -  &  - \\
4097 & 5350 & 3477 & 631 & 0.067 & 0.010 & 0.2769 & 2.04 &  -  &  - &  -  &  - \\
4446 & 5500 & 2880 & 851 & 0.078 & 0.006 & 0.0216 & 3.26 & 0.3240 & 2.29 &  -  &  - \\
5535 & 3900 & 2654 & 147 & 0.207 & 0.006 & 0.0000 & 11.53 & 0.0000 & 9.77 & 0.8491 & 2.42 \\
5548 & 3950 & 3424 & 211 & 0.086 & 0.021 & 0.1714 & 3.97 & 0.2901 & 2.86 &  -  &  - \\
5559 & 3950 & 3261 & 191 & 0.216 & 0.030 & 0.0221 & 5.20 & 0.6198 & 4.69 &  -  &  - \\
6060 & 4300 & 3791 & 272 & 0.155 & 0.045 & 0.1696 & 4.10 & 0.0642 & 7.92 & 0.4130 & 2.48 \\
6149 & 3950 & 3327 & 120 & 0.259 & 0.028 & 0.0231 & 7.86 & 0.4926 & 3.13 &  -  &  - \\
6259 & 4750 & 3337 & 431 & 0.104 & 0.009 & 0.0029 & 4.82 & 0.0000 & 5.59 &  -  &  - \\
6315 & 3950 & 3186 & 295 & 0.154 & 0.032 & 0.3825 & 2.27 &  -  &  - &  -  &  - \\
6861 & 4300 & 3255 & 290 & 0.132 & 0.015 & 0.0535 & 4.66 & 0.0301 & 3.30 &  -  &  - \\
6929 & 3900 & 3205 & 130 & 0.164 & 0.016 & 0.0005 & 7.27 & 0.1600 & 4.38 &  -  &  - \\
7181 & 4000 & 3072 & 143 & 0.175 & 0.010 & 0.0000 & 8.94 & 0.0000 & 6.55 &  -  &  - \\
7422 & 4350 & 2848 & 680 & 0.059 & 0.004 & 0.0763 & 3.32 & 0.2219 & 5.42 & 0.6330 & 1.84 \\
7632 & 3950 & 3218 & 77 & 0.290 & 0.013 & 0.0000 & 11.92 & 0.0000 & 12.13 &  -  &  - \\
8025* & 4000 & 3296 & 61 & 0.430 & 0.017 & 0.0000 & 15.61 & 0.2482 & 12.12 &  -  &  - \\
8038* & 4000 & 3391 & 187 & 0.150 & 0.025 & 0.1222 & 4.86 & 0.3512 & 3.33 &  -  &  - \\
8249 & 3950 & 3414 & 210 & 0.121 & 0.027 & 0.2193 & 3.47 & 0.0066 & 2.10 &  -  &  - \\
9267* & 5000 & 2896 & 253 & 0.122 & 0.003 & 0.0000 & 13.36 & 0.0000 & 28.30 & 0.0000 & 7.34 \\
9321 & 4250 & 3482 & 107 & 0.241 & 0.018 & 0.0521 & 10.48 & 0.3371 & 14.61 & 0.0000 & 3.22 \\
9961* & 4000 & 2409 & 396 & 0.053 & 0.006 & 0.0166 & 2.13 & 0.2968 & 1.59 &  -  &  - \\
\hline
\multicolumn{12}{|l|}{Ambiguous source in HS:CS$_{\{ V \}}$} \\ \hline
6813 & 3950 & 3604 & 114 & 0.248 & 0.060 & 0.4987 & 4.06 & 0.9985 & 4.64 &  -  &  - \\
7472 & 4300 & 3457 & 218 & 0.170 & 0.021 & 0.4229 & 4.67 & 0.2488 & 4.59 &  -  &  - \\
\hline
\multicolumn{12}{|l|}{Potential AA~Tau contaminant} \\ \hline
4766 & 4100 & 2774 & 59 & 0.4996 & 0.0003 & 0.0002 & 14.54 & 1.0000 & 13.35 &  -  &  - \\
6337 & 3950 & 2007 & 4 & 0.2412 & 0.0040 & 0.0001 & 11.15 & 0.0000 & 5.73 &  -  &  - \\
\hline
\end{tabular}
\end{table*}

\section{Spot properties of the IC~5070 YSO sample} \label{results}

\subsection{Object categorisation}

\begin{figure}
\centering
\includegraphics[width=0.99\columnwidth]{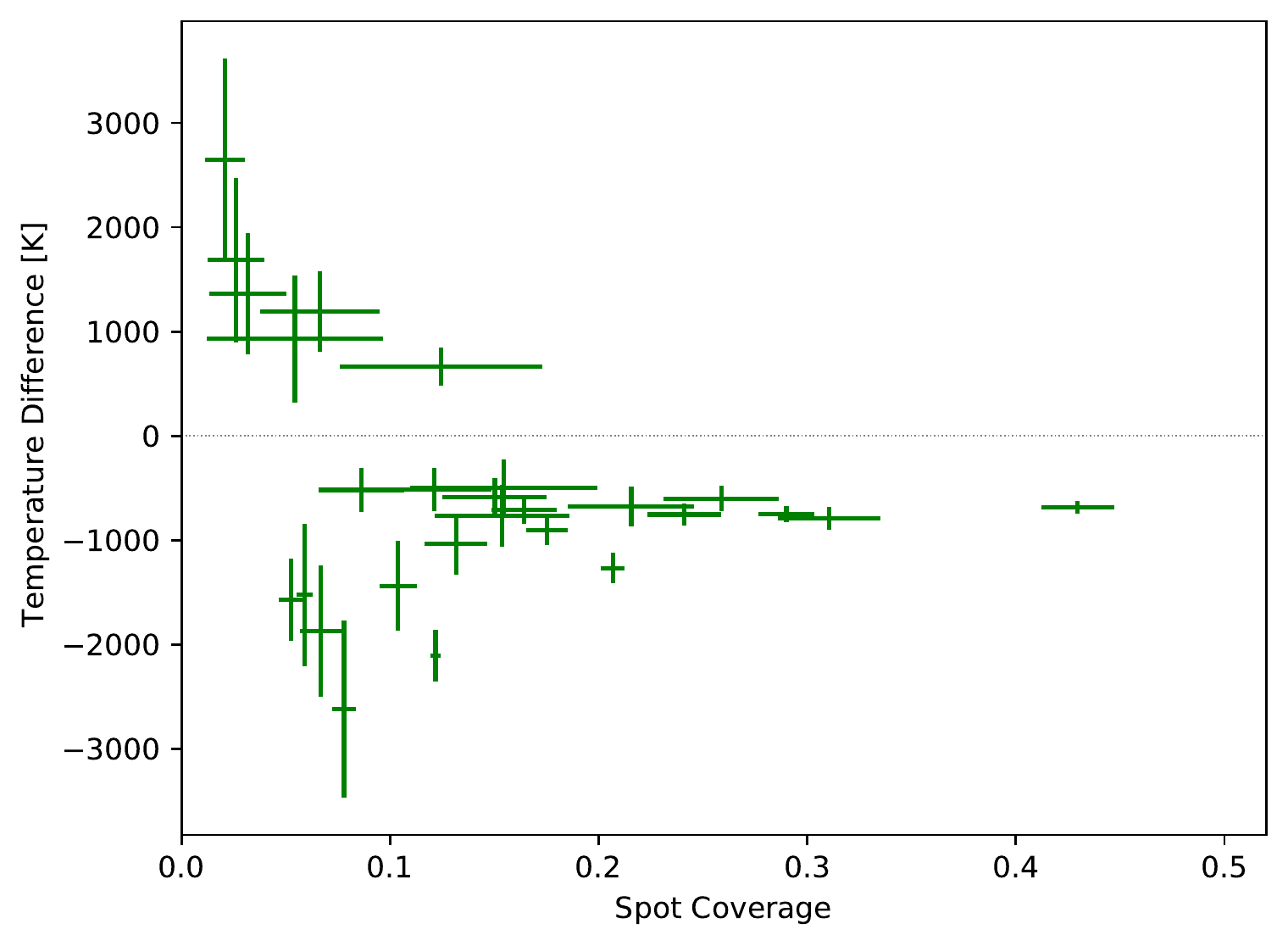}
\caption{ The determined spot properties of our YSO sample listed in Table~\ref{tbl_keynumbers} using $\hat{A}^o_{ \{ V \}}$. We plot the coverage of the visible surface and the temperature difference $T_S - T_\star$. The horizontal dotted line separates hot spots (top) from cold spots (bottom). The error bars represent the MAD uncertainty (see text for details). \label{fig_mov}}
\end{figure}

Based on the methodology outlined in the previous section, we have determined the best fitting spot properties for all objects, using the $\{V\}$ amplitudes.  We remind the reader that $\{V\}$ refers to set of amplitudes in $V$, $R$, and $I$ following the notation explained in Sect.~\ref{sec_anoteonnotation}. The spot temperatures and their MAD uncertainties are listed in Table\,\ref{tbl_keynumbers}. In the table we sort the objects into four categories: hot spots, cold spots, ambiguous sources (according to the HS:CS$_{\{V\}}$ ratio), and potential AA~Tau contaminants. There are six stars with hot spots, 21 with cold spots, two ambiguous object, and two potential AA~Tau contaminants. In Fig.\,\ref{fig_mov} we plot the spot coverage vs. the temperature difference of spot and star (with the MAD uncertainties) for the 27 stars with hot or cold spots. If the data are available, we have repeated the spot property calculations for $\{B\}$ (24 sources) and $\{U\}$ (5 sources). These values are not listed in Table\,\ref{tbl_keynumbers}.

Our classification of sources into hot/cold spot objects has been based on the $\{V\}$ data, because it is available for all objects. In Table\,\ref{tbl_keynumbers} we list the HS:CS ratios for all filter sets investigated. We also list the minimum signal-to-noise ratio (SNR) of any of the amplitudes in $\{V\}$, as well as the SNRs of the $B$-band and $U$-Band amplitudes, if available. An inspection of these values shows that the objects we consider ambiguous (0.4\,$<$\,HS:CS\,$<$\,0.6) are typically associated with SNRs lower than five. Note that low SNRs do not necessarily result in ambiguous sources. Amongst the stars with cold spots, there is only one source (6315) with a HS:CS ratio close to our adopted borderline. However, for two of the stars with hot spots (5886, 6393) the HS:CS ratio is just above the threshold, and two further objects (5575, 7954) are close to the threshold as well. In all those cases the SNRs of the amplitudes are very low. Thus, if one adopts a different cut-off for HS:CS, the number of hot spot objects in our sample is potentially decreased significantly. This discussion highlights that the distinction between hot and cold spots is only reliable when sufficient SNR amplitudes in multiple filters are available.

We find that there are three stars (5535, 5559, 7422) that are classified as having cold spots based on the $\{V\}$ data, which change to a hot spot when the $B$ or $U$ amplitudes are included into the analysis. Again, the change is caused by including amplitudes with a SNR below five. Similarly there are three objects (7465, 6060, 6149) that change from their cold spot classification to ambiguous, when shorter wavelengths amplitudes are included in the analysis. All of them have low SNR values for the $B$ and $U$ amplitudes. 

There are two stars in our sample that are potential AA~Tau contaminants. Objects 4766 and 6337 have spot solutions at the edges of the parameter space. In the case of 6337 the solutions are in exactly the place that is predicted by our simulations of variability due to gray extinction with an amplitude of approximately 0.3\,mag. Thus, we conclude that this source is an AA~Tau type object, with large dust grains in the inner disk warp. Object 4766 has much higher amplitudes which are similar but not the same in all the filters. The spot solution places it at a coverage of 0.5. Hence, we cannot fully exclude that this is a star with very large spots. However, most likely this is also an AA~Tau like source. There are two stars with ambiguous solution based on the $\{V\}$ data, which we are therefore unable to categorise. As indicated above, they generally have amplitudes with low SNRs. Both sources have additional $B$-band amplitudes available, which would classify one (6813) as hot and one (7472) as cold spot. However, again the SNRs of the amplitudes are below five.  

Our analysis has hence identified 21 objects with cold spots and six objects with hot spots. If amplitudes with SNRs of less than five are used in the analysis then objects can potentially not be characterised accurately. However, there are multiple cases with accurate characterisation despite low SNR amplitude. Thus, we refrain from excluding all objects with SNRs of the amplitudes below five. This would half the cold spot sample and only leave one hot spot source (7465). 

A majority of our stars have disks (see Section \ref{spots_stellar}). It may therefore seem surprising to find only so few objects with hot spots induced by accretion from a disk. There are two potential reasons for the scarcity of hot spots in our sample, one is spot instability, the other the properties of the underlying magnetic field.

Only stable accretion columns on stars are able to create periodic variability. Class\,II sources with unstable accretion do not produce regular periodic light curves that can be identified in the photometry \citep{2014EPJWC..6404004K}. We recall that the dataset used in the identification of our sample, was taken over a duration of 80\,d. Therefore, the accretion column is required to be somewhat stable for this period in order to have been included in our analysis. Thus, our study demonstrates that potentially only a small number of accreting stars is found in the stable regime of accretion, in the context of those models.

The second explanation for finding only few stars with hot spots is the magnetic field configuration. Pre-main sequence stars that are fully convective have strong, simple magnetic fields. With the development of a radiative core the dipole becomes weaker and the field becomes more complex \citep{2012ApJ...755...97G, 2011MNRAS.417..472D}. \citet{2015MNRAS.454.4037T} show that the timescale of developing a radiative core depends on the mass ranging from 0.5~Myr (for 2~M$_\odot$) to 9.3~Myr (for 0.5~M$_\odot$). Based on the Hertzsprung-Russel diagram for the IC\,5070 region discussed in \citet{2021MNRAS.506.5989F} and $T_{\rm eff}$ values, the masses of our objects should be in the range of 0.5 to 2 M$_\odot$. In combination with the typical age of 1~Myr,  our sample therefore represents a mix of simple and complex magnetic fields. While complex fields would result in hot spots at a range of latitudes, simple fields give preferably spots at high latitudes which are less likely to give rise to photometric variability.

\subsection{Spot properties}\label{sec_IC5070targets}

The detected hot spots have temperatures from about 700\,K to 2800\,K above the stellar surface temperature. The coverage ranges from 0.02 to 0.12. There is a general trend that larger temperature differences are associated with smaller coverage. Note that the trend basically follows the detection limit of our objects, as smaller spots with less temperature difference create lower amplitude variations, which are more likely not to be included in our sample. However, it is clear that there are no large hot spots covering more than 10~\% of a hemisphere with a  temperature difference larger than 1000\,K. Generally, the spot temperatures are consistent with low temperature contrasts on low mass stars and brown dwarfs observed by e.g. \citet{1993A&A...272..176B, 1995A&A...299...89B, 2009MNRAS.398..873S, 2012MNRAS.419.1271S, 2016MNRAS.458.3118B} and predicted by \citet{1998ApJ...492..743M, 2006MNRAS.370..580K}. We note that our simplified model determines the spot temperature averaged over the spot coverage, therefore, the maximum spot temperatures on the surface could be significantly higher.

The general trends of the detected cold spots mirror what we have seen for the hot spots. Higher temperature differences are typically associated with smaller spots. We do not find stars with large spots that also have a large temperature contrast. The sample contains objects with star-spot temperature difference between 500\,K and 2500\,K, with most being between 500 and 1000\,K. Smaller temperature differences than 500\,K do not create large enough amplitudes and such objects are thus not included in the sample. The two stars with highest star-spot temperature difference (4097, 4467) are also the two with the highest effective temperatures, around 5500\,K. The absolute spot temperatures for these two objects are in line with the other stars that have effective temperatures of the order of 4000\,K. 

The cold spot coverage ranges from 0.05 to 0.43, with typical values between 0.1 and 0.3. The maximum coverage is well constrained, while the lower bound is a bias in the sample. With the typical stellar surface temperatures and the wavelength range of our observations, cold spots with smaller coverage do not cause large enough variability. Compared to the hot spots, the coverage for cold spots can be much larger, and reach up to half the visible surface. This has also been seen in previous works, see, for example, \citet{1992ASPC...34...39S, 1995A&A...299...89B}.  Similar to the hot spots, we note that the temperatures are averages over the spot area.

\begin{figure*}
\centering
\includegraphics[width=0.99\columnwidth]{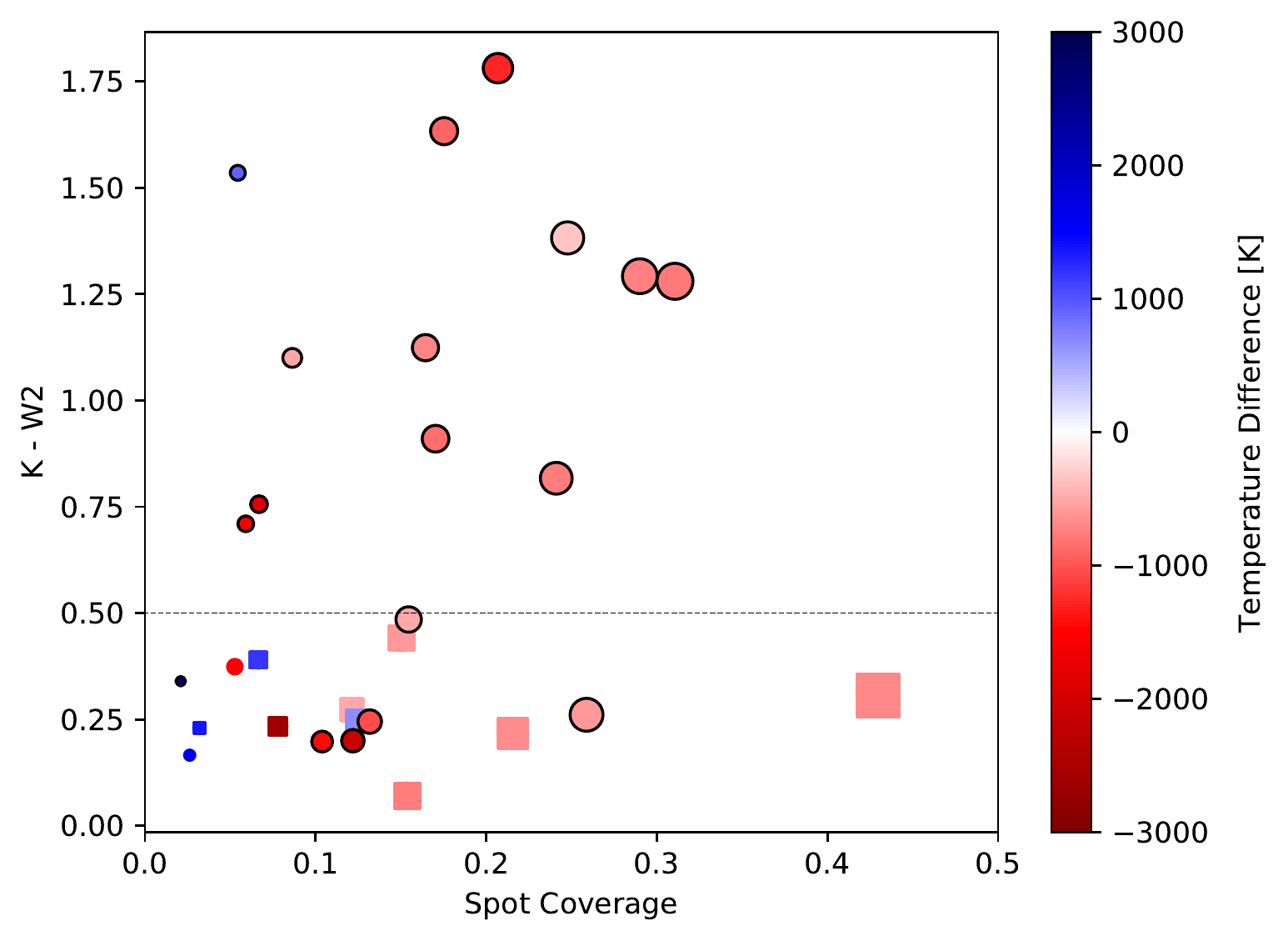} \hfill
\includegraphics[width=0.99\columnwidth]{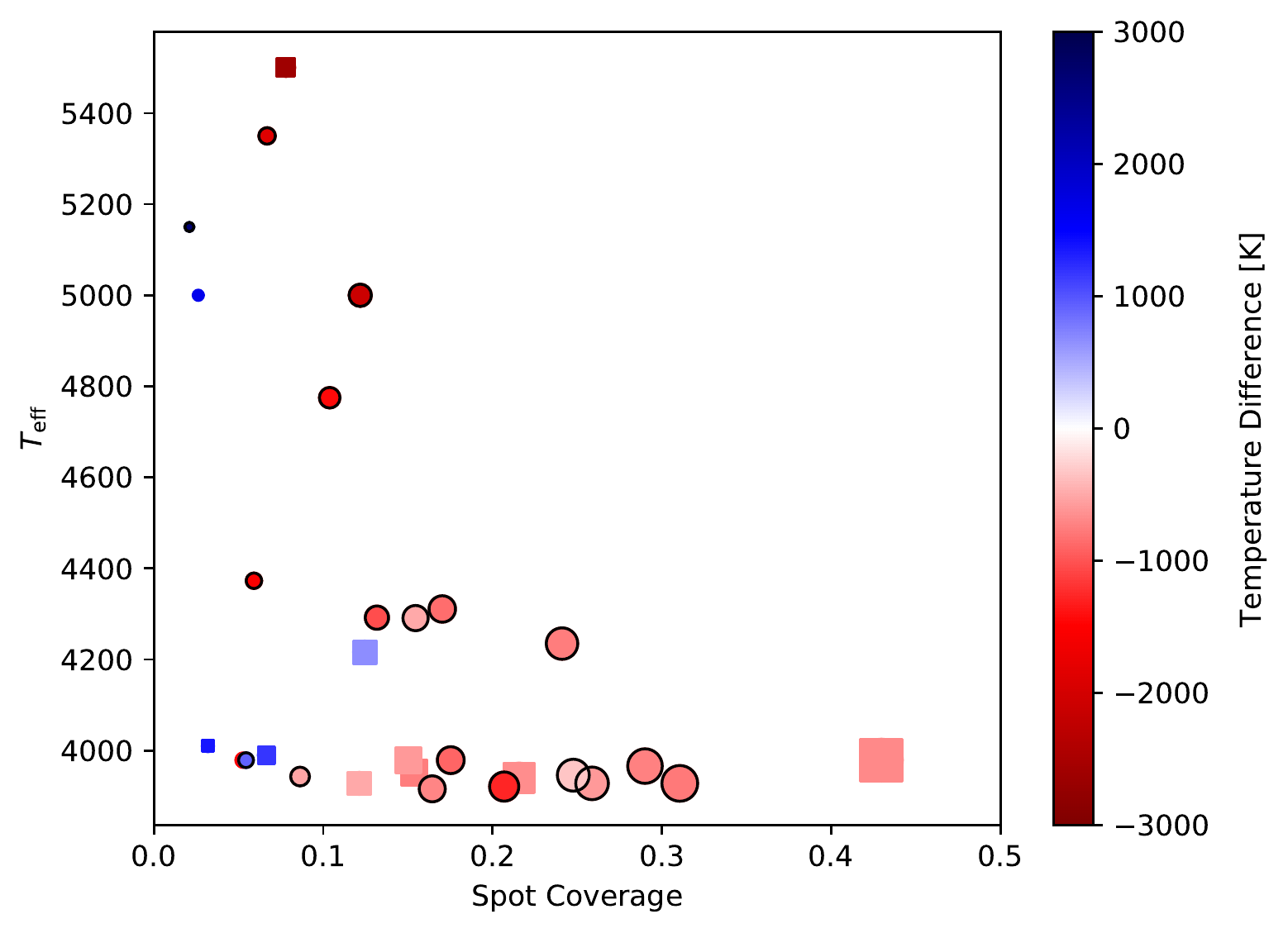}
\caption{  {\bf Left:} Spot coverage against $K-W2$ colour for our objects. The marker size represents the spot coverage and the colour the temperature difference $T_S - T_\star$. Hot spots are blue and cold spots are red. The horizontal dashed line separates sources with and without disk excess emission. Objects classified as Class\,II according to $\alpha _{SED}$ are highlighted with a black circle. Square symbols indicated sources with SNR\,$<5$ in $W3$ or $W4$. {\bf Right:} Spot coverage vs. stellar effective temperature. The colour coding and symbol size is the same as in the left panel. \label{fig_stageanddisk}}
\end{figure*}

\subsection{Spots and stellar properties}
\label{spots_stellar}

In this section we aim to discuss if and how the evolutionary stage or stellar properties of the YSOs in our sample influence the spot properties. It is worth re-iterating here that the original sample of young stars used has been obtained solely by the presence of periodic variability and their astrometric properties \citep{2021MNRAS.506.5989F}. In other words, all objects have parallax and proper motion values that indicate they are members of the IC\,5070 star forming region. The typical age of members of this region is 1 Myr \citep{2020ApJ...899..128K}.

First, we check for the presence of disks, using indicators for the infrared excess. Traditionally the slope ($\alpha$) of the infrared spectral energy distribution (SED) has been used to distinguish between stars with (Class~II) and without disks (Class~III) \citep{1987IAUS..115....1L, 1993ApJ...413L..47M}. \citet{2013Ap&SS.344..175M} has used Eq.\,\ref{eq_SED} to determine the SED slope from WISE photometry:  

\begin{equation}
    \alpha _{SED} = 0.36(W1 - W2) + 0.58(W2 - W3) +0.41(W3 - W4) - 2.90 
    \label{eq_SED}
\end{equation}

A value for $\alpha _{SED}$ between zero and  $-1.6$ indicates Class\,II, and $\alpha_{SED} < -1.6$  Class\,III.  Based on the $\alpha_{SED}$ values the majority of our sample is Class\,II -- 25/31 or 80\%. However, this metric is only deemed reliable if the star has SNR\,$ > 5$ in all four WISE bands \citep{2013Ap&SS.344..175M}. Nine of our stars do not meet this standard in $W4$. Five of these nine also fail the SNR requirement in $W3$. Most of the ones failing the SNR criterion (5/9) are nominally Class\,III, the remaining 4 Class\,II. Thus, using $\alpha_{SED}$ we can reliably classify 21/31 or $68$\% as Class\,II. We note that two of these are the AA~Tau contaminants identified in this paper, confirming that they host a circumstellar disk.

An alternative and simpler disk indicator is the $K - W2$ colour, which traces again the slope of the SED, but without using the two longest wavelengths in WISE. For a discussion about the relation between mid-infrared colours and slope of the SED, see \citet{2020A&A...642A..86T}. In $K-W2$, values above $\sim 0.5$\,mag generally indicate an infrared excess due to disk material. Note that the $K-W2$ excess is potentially influenced by long term variability of the star due to the time delay of several years between 2MASS and WISE. According to this criterion 12/31 stars have infrared excess due to a disk, fewer than when using the slope of the SED, as above. It is worth noting here that $K-W2$ only traces material in the inner disk, within 1\,AU. Thus, it will not identify more evolved disks, with an AU-scale inner hole or a partially depleted inner region. Taken together both criteria, we conclude that about two thirds of our sample are objects hosting a circumstellar disk (Class~II), and about half of those are likely to be evolved.

In the left hand panel of Fig.\,\ref{fig_stageanddisk} we show the spot coverage against the $K-W2$ disk excess indicator. The horizontal dotted line separates the objects with and without disk excess. Class~II objects according to $\alpha_{SED}$ are marked with an extra black circle; squares denote stars where the SNR is too low to distinguish between Class~II and Class~III. The symbol size is proportional to the spot coverage and the symbol colour shows the temperature difference between star and spot based on the legend. 

There is no statistically significant correlation of the spot properties with the indicators for the presence of the disk. For the stars with cold spots the coverage and temperature difference of spot and star are homogeneously distributed amongst the objects with and without excess emission. Furthermore, the small number of hot spot sources also does not show any preference to occur on sources with infrared excess. Maybe most noteworthy is the presence of hot spots on stars that do not show infrared excess in $K-W2$. As pointed out above, the lack of $K-W2$ excess does not necessarily exclude the presence of a disk, or the presence of accretion. A good example is the star TW~Hya: it has a $K-W2$ colour of 0.4\,mag, but hosts an active accretion disk, with a large inner cavity \citep[e.g.][]{2006ApJ...648.1206J}. We note that most of the stars with hot spots without $K-W2$ excess also do not have sufficient SNR to identify the disk from $\alpha_{SED}$. Apart from a disk with an inner cavity, an alternative explanation for hot spots on stars without apparent infrared excess is emission from chromospheric plage, as for the star V1598\,Cyg found in \citet{2020MNRAS.497.4602F}. This could be akin to plages or faculae on the Sun, but for much more active stars. As a reminder, with our methodology, if stars have dark spots and plage, we would only detect hot spots if the plage dominates. \citet{2016A&A...589A..46S} investigated if a broadband light curve for the Sun would be dominated by cool or hot regions, and the answer is it depends on viewing angle and timescale. This illustrates that even in the Sun hot spots can show up in light curves.
 
Similarly to the evolutionary stage indicators, there is no significant correlation of the spot properties with the stellar rotation period. Tentatively, there is a link between spot coverage and effective temperature. These two quantities are plotted in the right panel of Fig.\,\ref{fig_stageanddisk}. We can see that the maximum coverage decreases with increasing $T_{\rm eff}$. This indicates that cooler, lower mass stars can have larger cold spots. This could be caused by these stars having deeper convection zones, because they are too low mass or too young to have developed a radiative zone near their core. Because our sample contains only a small number of stars with high $T_{\rm eff}$, we note that we cannot exclude the possibility of an observational bias causing this trend.  If most spots are small, then randomly drawing coverage values for a few high temperature objects and a large number of low temperature sources, will create a similar trend. However, the median and maximum spot coverage clearly decreases with effective temperature.

\section{Conclusions}

We have investigated a sample of 40 YSOs that were selected based on their periodic light curves and astrometric properties as members of the Pelican Nebula star forming region by \citet{2021MNRAS.506.5989F}. We have selected 31 of these sources with photometry available in at least three optical filters ($V, R,$ and $I$) during an 80\,d period in the summer 2018, provided by the HOYS project \citep{2018MNRAS.478.5091F}. On average, each source has about 270 photometric data points in each of the filters. The stars in our sample are a mix of Class~II and Class~III sources, covering a range of effective temperature from 3900 to 5500~K.

A simple spot model has been developed. It assumes a single temperature dominating spot, situated on a homogeneous stellar surface without limb darkening. Using stellar atmosphere template spectra we model peak-to-peak amplitudes for a range of spot temperatures and surface coverages. These are compared to the measurements for our sources in at least three optical filters to evaluate the best fitting spot properties. Measurement uncertainties are used to conduct a full error propagation. The typical uncertainties in spot temperatures are of the order of 200\,K. The spot coverage can be evaluated within a few percent of the projected stellar surface area. We also find that systematic uncertainties caused either by uncertain stellar properties (e.g. $T_{\rm eff}$, $\log(g)$, [M/H]) or the choice of filter, are smaller than the statistical errors of our method. Furthermore, our method allows us to identify AA~Tau type contaminants in the sample of periodic YSO variables.

In our sample for two thirds (21/31) of the stars the variability is caused by cold spots, 19\,\% (6) show evidence for hot spots and two are AA~Tau contaminants. A further two objects are considered ambiguous and cannot be classified. For these two stars, the signal to noise in the amplitude measurements is very low. The identified hot spots have a coverage of less than 0.15 and a temperature up to 3000\,K above the surface of the star.  This is in agreement with previously published observational and theoretical works. The cold spots have a temperature of up to 2500\,K colder than the stellar surface, and the spots can cover up to 0.4 of the visible surface. The limiting temperature contrast required for spots to be identified and included in our sample is 500\,K. Cold spots also need to cover more than 0.05 of the stellar surface to be detectable in the HOYS photometry. 

In our sample, large cold spots are typically found on relatively cool stars with $T_{\rm eff} < 4500$\,K. This could be a result of the deeper convection zones in these objects. A larger sample of sources as well as a study of the time evolution of the cold spots are required to verify this. Our long-term HOYS data will be used in a future paper to investigate the evolution of the spots. 

The spot properties do not show any significant correlation with the stellar rotation period or the presence of a disk. In particular, we find that hot spots are present on objects without any significant infrared excess emission. Thus, perhaps counter intuitively, hot spots can be present on sources without a detectable inner disk. This indicates the possibility of accretion across an inner disk cavity or the presence of plage.

The small number of detected hot spots in our sample of mostly Class\,II sources can be explained in two ways. The YSOs are dominated by objects with unstable accretion over the 80\,d period, which will not create an identifiable periodic light curve. It is also possible that we are missing stars with hot spots found mostly at high latitudes. Unbiased samples with similar disk frequency are expected to contain a larger fraction of stars with hot spots.

\section*{Acknowledgements}

We would like to thank all contributors of observational data for their efforts towards the success of the HOYS project. This work was supported by the Science and Technology Facilities Council.



\section*{Data Availability Statement}

The data underlying this article are available in the HOYS database at http://astro.kent.ac.uk/HOYS-CAPS/.


\bibliographystyle{mnras}
\bibliography{bibliography} 


\clearpage
\newpage

\appendix

\section{Systematic uncertainties of the Spot properties} \label{sec_systematics}

Here we detail the analysis of systematic uncertainties of our methodology. 

\subsection{Choice of stellar properties}

The modelled amplitudes $\hat{A}^m_{ \{ \lambda \}}$ are all calculated from one stellar atmosphere model, with a fixed stellar temperature, metallicity, and surface gravity. The assumption of solar metallicity and $\log(g) = 4.0$ is reasonable but not measured for all YSOs. As discussed in Sect. \ref{sec_data_Tref} the stellar temperature is not known for all the YSOs and in those cases has been set to the sample median. Hence, the effects of altering these stellar parameters in the models was investigated individually to quantify their influence on the results. 

We use a {\it simulated} spot for which we generated a set of $\hat{A}^s_{ \{ \lambda \}}$ amplitudes. They were generated for a star with $T_\star = 4000$\,K, $\log(g)=4$, [M/H]\,=\,0.0 and for spot properties $T_S = 3200$\,K and $f = 0.1$. These correspond to typical values for our sample of YSOs, as shown in Sect.~\ref{sec_IC5070targets}.

Alternative sets of $\hat{A}^m_{ \{ \lambda \}}$ were generated with one of the parameters ( $T_\star , \log(g),$ [M/H]) adjusted and the rest kept at the original value. We varied $\log(g)$ between $3.5\,\leq \log(g) \leq \, 4.5$, and the metallicity between $-0.5\,\leq $\,[M/H]\,$\leq \, 0.5$, both in steps of 0.5. The stellar temperature was varied from 3800\,K to 4200\,K in 100\,K steps. 

We investigate the effect on the peak-to-peak amplitudes when the metallicity and temperature were altered. The changes are more significant as the wavelength decreased and hence $\hat{A}^m_{U}$ is the most changed. However the amplitudes in all bands underwent some change. Altering $\log(g)$ affected all $\hat{A}^m_{ \{ V\}} $ equally and to a lesser extent than $\hat{A}^m_B$ and $\hat{A}^m_U$. This was the case for both the ATLAS9 and PHOENIX atmosphere models.  

\begin{figure}
\centering
\includegraphics[width=0.99\columnwidth]{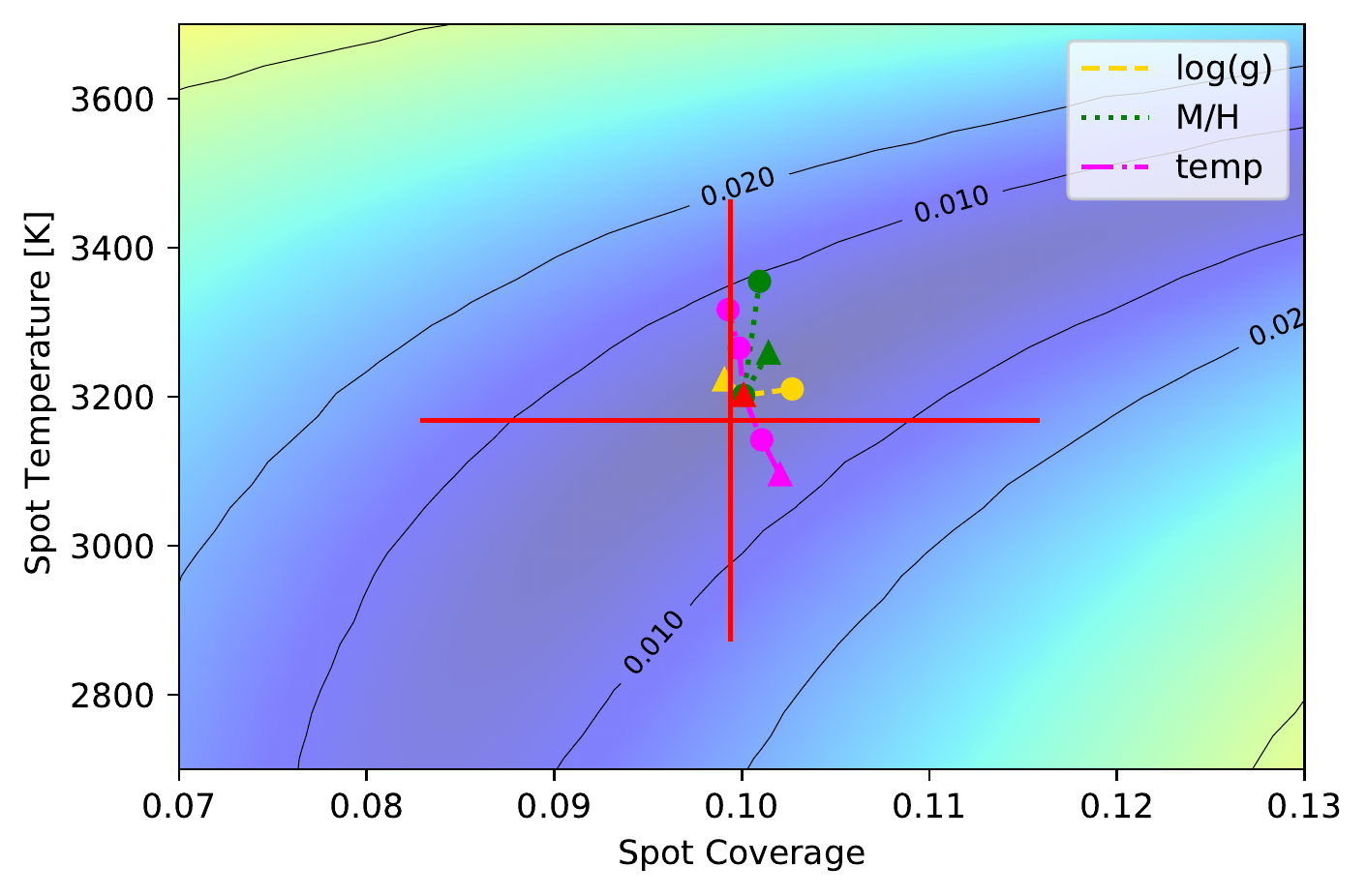} \\
\includegraphics[width=0.99\columnwidth]{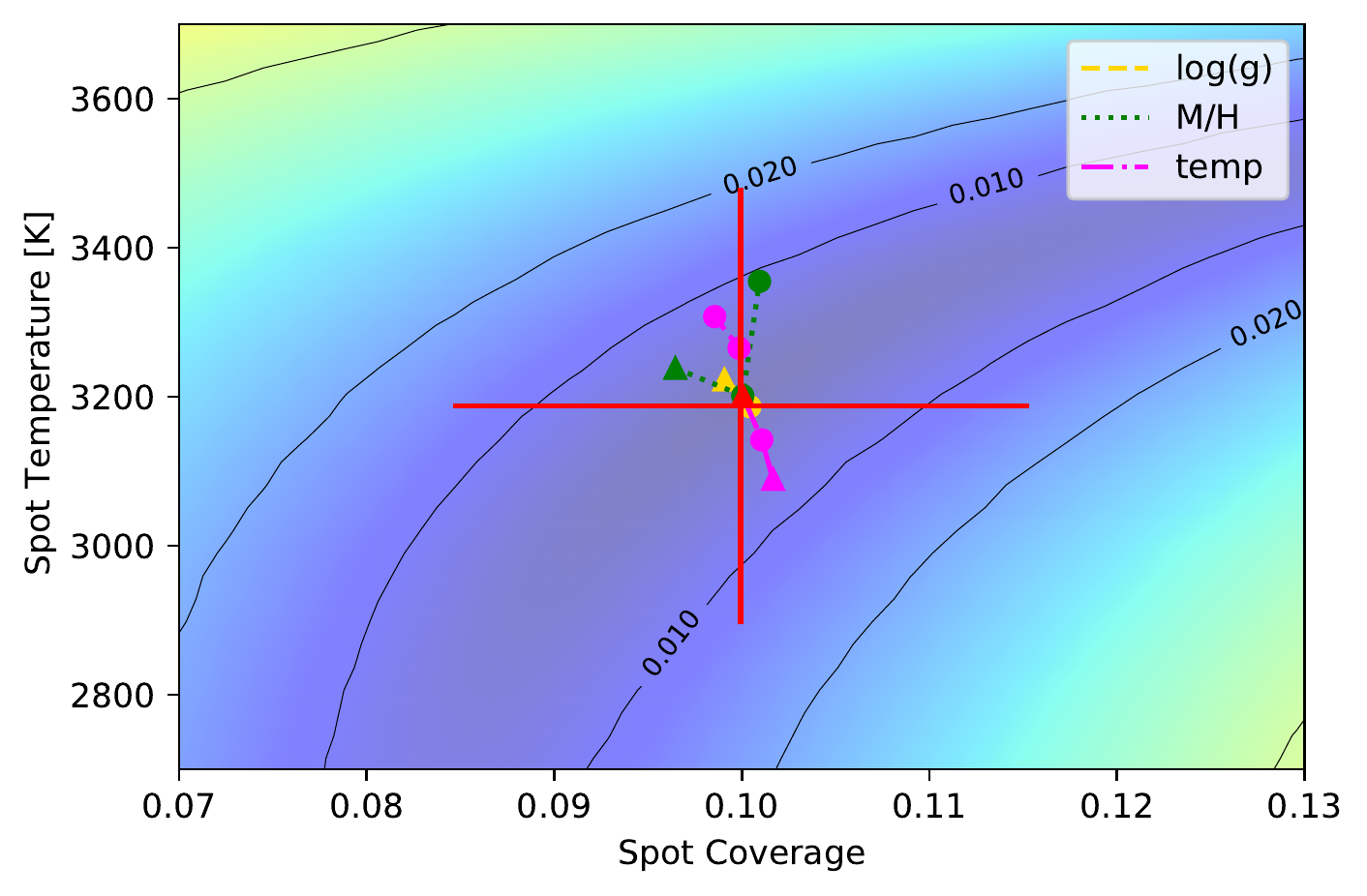} \\
\includegraphics[width=0.99\columnwidth]{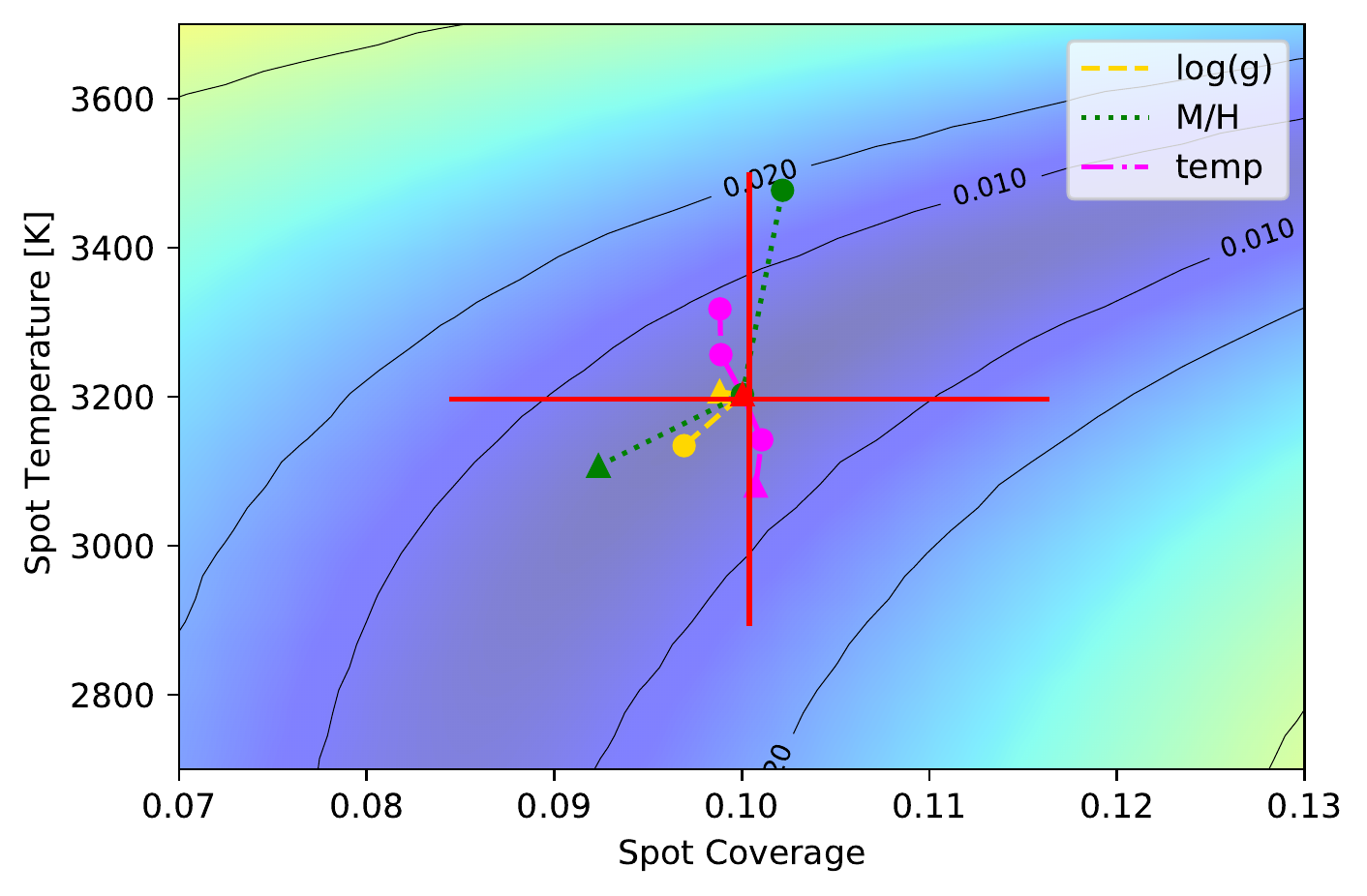}
\caption{ {\bf Top:} Contour plot showing $RMS_{\{V\}}$ between $\hat{A}^s_{ \{ V \}}$ and  $\hat{A}^m_{ \{ V \}}$ for a simulated spot with parameters $T_\star = 4000 \rm{K}$, $T_S = 3200 \rm{K}$ and $f = 0.1$. The red triangle indicates the best fitting model with stellar parameters $T_\star = 4000~\rm{K}, \log(g) = 4.0$, and [M/H]~=~0.0. The pink dash-dot track indicates the best models when $3800~\rm{K}\,\leq T_{\star} \leq \, 4200~\rm{K}$, in 100~K steps. The green dotted track shows the best models when  $-0.5\,\leq $\,[M/H]\,$\leq \, 0.5$, in 0.5 steps. The yellow dashed track shows the best models when $3.5\,\leq \log\,(g) \leq \, 4.5$, in 0.5 steps. The start of each track is marked by a triangle. The red cross indicates the MAD uncertainty - see text for details. {\bf Middle:} As top panel but for $\hat{A}^s_{ \{ B \}}$. {\bf Bottom:} As top panel but for $\hat{A}^s_{ \{ U \}}$. \label{fig_k210_newtracks} }
\end{figure}

In the top panel of Fig.~\ref{fig_k210_newtracks} we show the inferred spot parameters when we use the $\hat{A}^m_{ \{ V \}}$ set created with the different stellar parameters. The background contour in the figure is $RMS_{\{V\}}$ between $\hat{A}^s_{ \{ V \}}$ and $\hat{A}^m_{ \{ V \}}$ with the original stellar parameters. Following the procedure laid out in Sect.\,\ref{sec_statistical}, we determine the best fitting spot parameters and the MAD uncertainties. For the latter we use the mean uncertainties of the amplitudes of the YSO sample for the $\hat{A}^s_{ \{ \lambda \}}$ values.

We show the position of the best fitting model for each of the variations in stellar temperature, metallicity and $\log(g)$ discussed above. There is a small off-set of the best fitting model to the actual input spot parameters, which is caused by the resolution of the set of $\hat{A}^m_{ \{ \lambda \}}$, meaning there is no entry in  $\hat{A}^m_{ \{ \lambda \}}$ generated for exactly the input spot properties.

We can see that the best fitting models for all used stellar parameters fit within the MAD uncertainties for the spot properties. As the stellar temperature increases the best model moves to a higher spot temperature position maintaining the same spot coverage. The total range of stellar temperatures covered is 500~K, and this covers less than half of the MAD uncertainty. This validates our decision to fix the stellar temperatures to the nearest 50~K step, as the effect on the final results is minimal. Varying $\log(g)$ has almost no effect on the inferred spot properties. The metallicity comes the closest to straying outside the MAD uncertainty. However, the PHOENIX models for the metallicity [M/H]~=~0 were updated to a later generation of models in 2021. Until the models for other metallicites are updated, which is in progress (Diaz, priv. comm), we are not able to investigate this further. The temperature and $\log(g)$ tracks all use [M/H]~=~0, and as such use the updated models.

As discussed above, we require a minimum of three amplitudes for inclusion in this work. Hence, for all objects that we investigate, we have $\hat{A}^o_{ \{ V \}}$. But for some objects more data are available ($\hat{A}^o_{ \{B \}}$ and $\hat{A}^o_{ \{ U \}}$) to infer the spot properties. The recovered spot properties in the top panel of Fig.\,\ref{fig_k210_newtracks} are obtained using $\hat{A}^o_{ \{ V \}}$. The results are extremely similar when including one or both of the shorter wavelength filters. We show this in the middle and bottom panel of Fig.\,\ref{fig_k210_newtracks}, where the inferred properties for the same simulated spot are displayed when using $\hat{A}^o_{ \{B \}}$ and $\hat{A}^o_{ \{ U \}}$. The effect of varying the stellar parameters is largely consistent between filter sets. The effect of changing $\log(g)$ is minimal regardless of filter choice. The effect of varying the stellar temperature leads to a wider spread in spot coverage and changing [M/H] has a similar effect as for just $\hat{A}^o_{ \{ V \}}$. However, it is important to note that in all cases the changes in spot properties remain significantly below the MAD uncertainty.  

There are some small systematic differences, however. The magnitude of the MAD uncertainty of the spot properties changes minimally with filter selection, and slightly  increases when more shorter wavelength filters are included. The median value of the spot temperature is slightly moving towards the stellar temperature and the spot coverage increases slightly when using the amplitudes in $B$ and $U$.

\begin{figure}
\centering
\includegraphics[width=0.99\columnwidth]{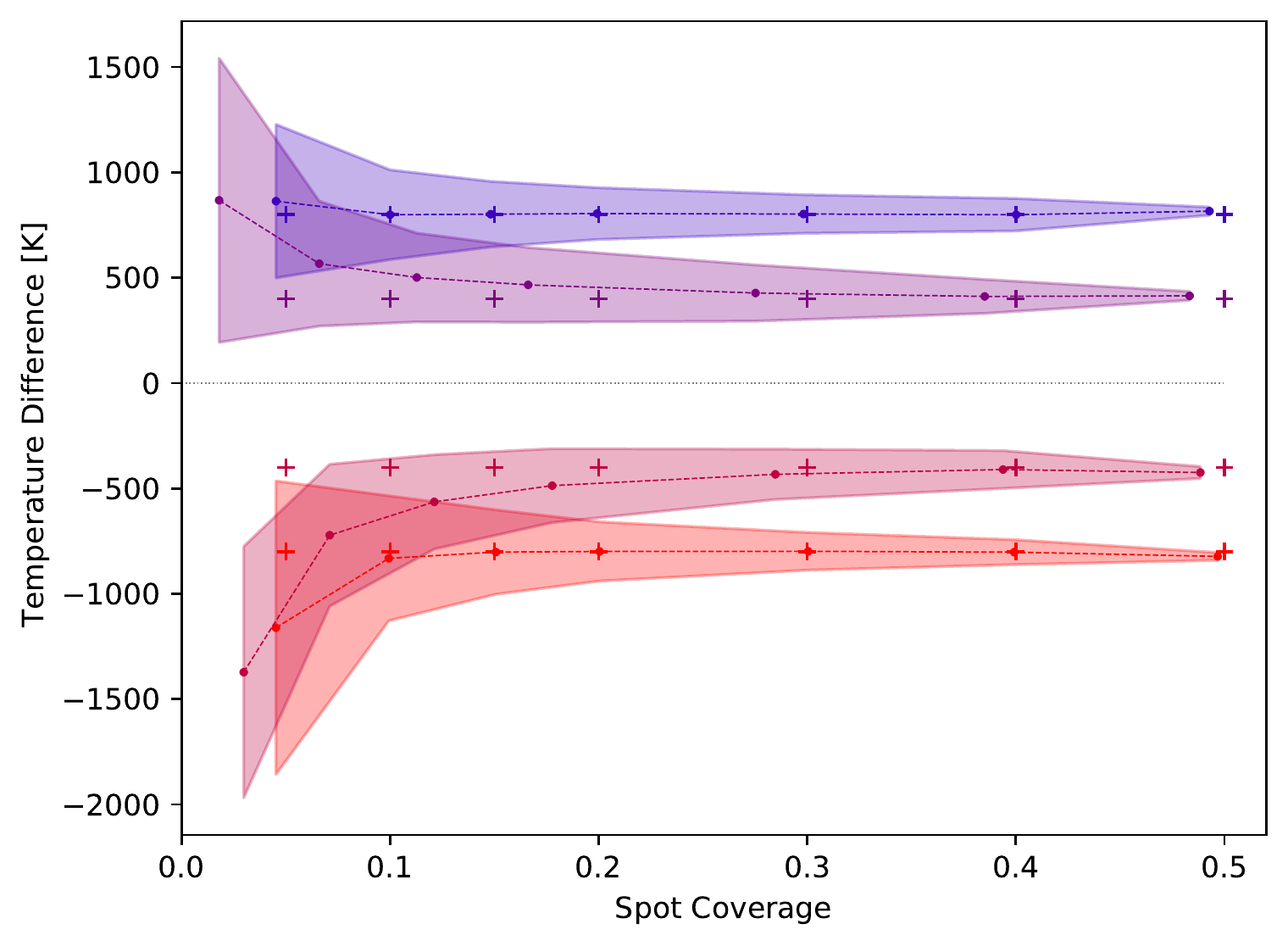} \\
\includegraphics[width=0.99\columnwidth]{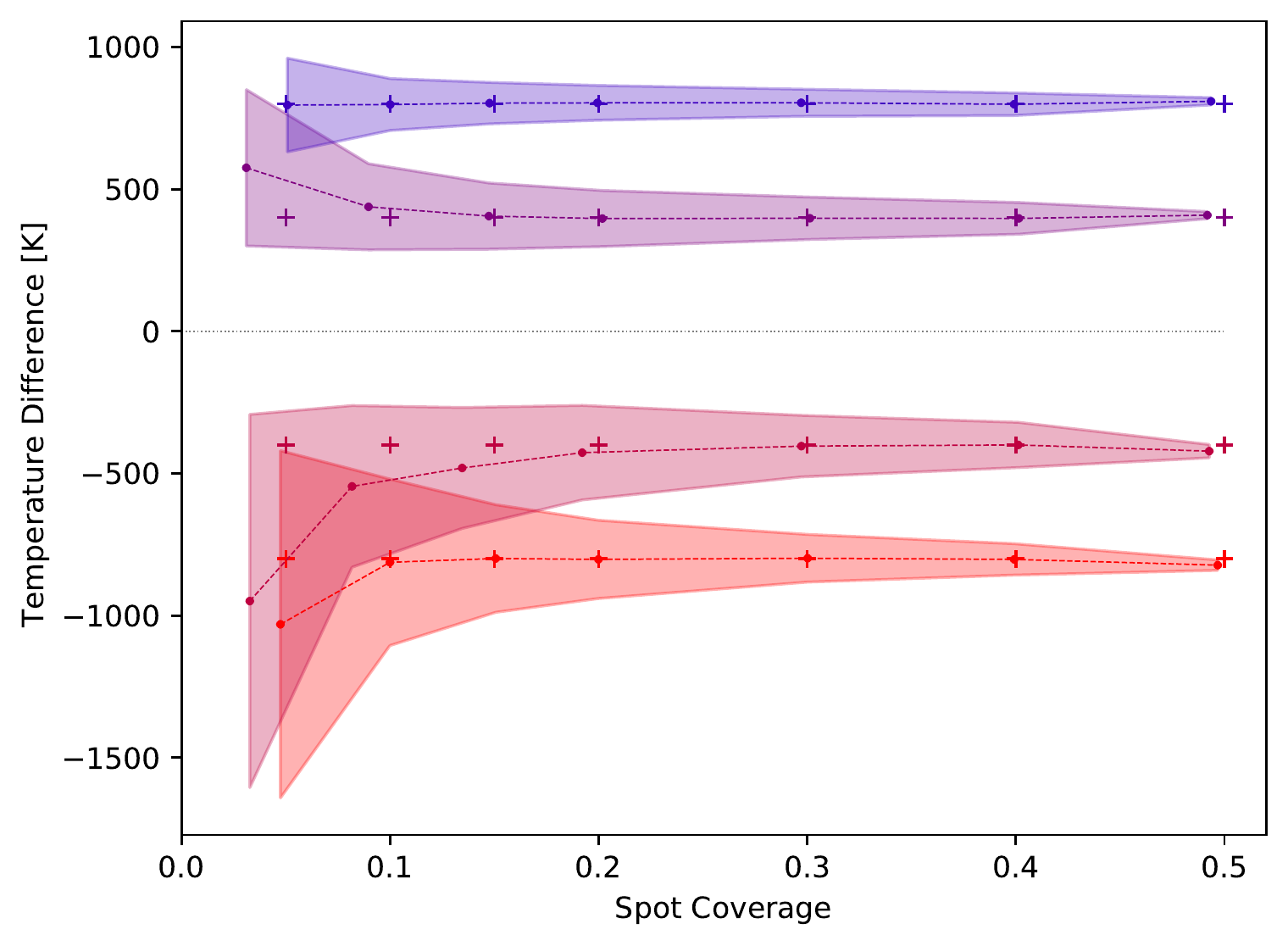} \\
\includegraphics[width=0.99\columnwidth]{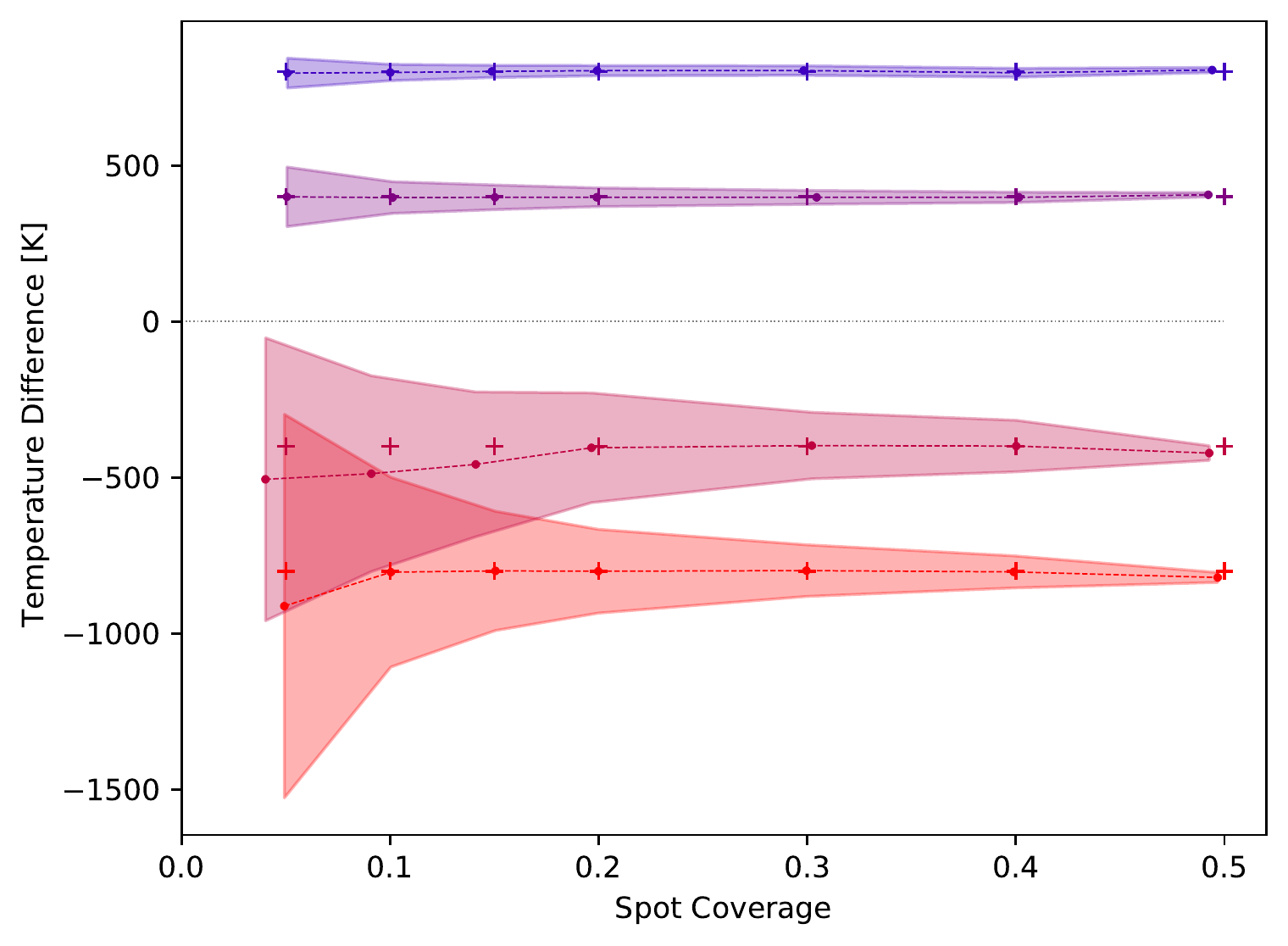}
\caption{ {\bf Top:} The recovered spot properties and uncertainty in temperature (shaded area) for a series of amplitudes $\hat{A}^s_{\{ V \}}$ for simulated spots. Crosses mark input spot values and circles the recovered properties. The dashed lines connect the recovered values for the same spot-star temperature difference. {\bf Middle: } As top panel but for $\hat{A}^s_{ \{ B \}}$. {\bf Bottom: } As top panel but for $\hat{A}^s_{ \{ U \}}$. \label{fig_known_v}}
\end{figure}

\subsection{Recovering simulated spot properties}\label{spot_simu}

We have found that the statistical uncertainties are always larger than the systematic uncertainties that can arise if we do not know the exact stellar parameters. However, we have seen that there are small systematic shifts, even if they are smaller than the statistical uncertainties. In this subsection we investigate how the size of these systematic shifts changes with spot properties and filters used.

Our sample median stellar temperature is 3979~K. Hence, for this test we use 4000\,K as well as $\log(g) = 4.0$ and [M/H]\,=\,0.0. We create amplitude sets $\hat{A}^s_{\{\lambda \}}$ for simulated spots of different properties. We use seven spot coverage values of $f= 0.05, 0.1, 0.15, 0.2, 0.3, 0.4, 0.5$. For each of those, two sets of hot and cold spots are generated. One has a temperature difference of 400\,K and one of 800\,K compared to the stellar temperature. For each of these 28 cases, we determine the spot properties and MAD uncertainties following the procedure laid out in Sect.~\ref{sec_statistical}. Again we use the average $\sigma \left( \hat{A}_\lambda^o \right)$  of our sample to generate $\hat{A}^v_\lambda$ for the spot property uncertainty calculation.

In the top panel of Fig.\,\ref{fig_known_v} we compare the recovered spot properties using amplitudes $\hat{A}^s_{\{ V \}}$, with the input spot properties. The crosses mark the input properties and circles the recovered values. The dashed lines connect the recovered spot temperatures for each of the four values of input temperature differences between spot and star. The shaded areas indicate the MAD uncertainty in spot temperature. The uncertainties in coverage are not shown but are typically of the order of 0.035. 

As one can see, the trend is to recover the spot as slightly smaller and further away from the stellar temperature compared to the input values. The spot temperature uncertainties increase systematically for smaller spot coverage values and also with decreasing temperature difference of spot and tar. This is caused by the fact that the amplitudes for the smallest spots have very low signal-to-noise ratios. The exception is the case for $f = 0.05, T_S = 3600~\rm{K}$, where the temperature values cluster around the 2000~K limit of the PHOENIX models, and thus artificially create small uncertainties. 

As demonstrated in Sect.~\ref{results}, we have not found any cold spots in our IC~5070 sample that have a coverage of $f < 0.05$, which could be mimicking the $f = 0.05, T_S = 3600~\rm{K}$ case. This is likely because the signal-to-noise ratio of the amplitudes are too low, and they hence have not been included in our sample We find that the minimum SNR of all amplitudes in the sample is greater than the SNR for the $f = 0.05, T_S = 3600~\rm{K}$ case. The largest modeled test spots have the greatest SNR and accurately recover the spot temperature but underestimate the coverage, typically by less that $0.05$.

\subsection{Choice of filters}

The recovery of simulated spots has been repeated using $\hat{A}^s_{\{B \}}$ and $\hat{A}^s_{\{U \}}$ to test the effect of including shorter wavelength filters. In the middle and bottom panel of Fig.~\ref{fig_known_v} we show the spot recovery for amplitudes generated for the same spot coverage and temperatures as in the top panel, but using $\hat{A}^s_{\{B \}}$ and $\hat{A}^s_{\{U \}}$. 

The inclusion of the shorter wavelength data reduces the uncertainties in temperature and coverage significantly for hot spots. The accuracy and systematic off-sets for the recovery of the cold spots is almost not changed. There is an apparent increase in the temperature uncertainty for the smallest spots, but this is solely due to the fact that now those models do not cluster at the edge of the parameter space anymore. This lack of improvement for the cold spots is understandable due to the typical temperatures of our YSOs and spots, whose spectral energy distribution peaks in the $V$, $R$, and $I$ filter range. As we have seen in Sect.\,\ref{results}, most of our objects show cold spots and only a fraction has $B$ and $U$ data (see Table\,\ref{tbl_ampe}). Thus, not having $B$ and $U$ data has no significant influence on the accuracy of our determination of the cold spot properties, which represent the vast majority of our sample.

\bsp	
\label{lastpage}
\end{document}